\documentclass[12pt]{elsarticle}
\usepackage{amssymb,amsmath}
\usepackage{mathrsfs}
\usepackage{fancyvrb}
\usepackage{hyperref}

\journal{\hspace{-10em}\colorbox{white}{\\
{\it NORDITA-2013-11} $\qquad$ {\it Imperial/TP/13/SL/01}}}

\makeatletter
\def\paragraph{\secdef{\els@aparagraph}{\els@bparagraph}}
\def\els@aparagraph[#1]#2{\elsparagraph[#1]{#2.}}
\def\els@bparagraph#1{\elsparagraph*{#1.}}

\makeatother

\def\appendixref{{}}

\def\a{{\alpha}}

\def\[{\left[}
\def\]{\right]}

\def\a{\alpha}

\def\<{\langle}
\def\>{\rangle}

\def\pint{-\hskip-0.41cm \int}
\def\i2{\frac{i}{2}}

\def\bq{{\bar q}}

\def\<{\langle}
\def\>{\rangle}

\def\cF{{\cal F}}

\def\cT{{\cal T}}

\def\pint{-\hskip-0.41cm \int}
\def\i2{\frac{i}{2}}

\def\Tr{{\rm Tr}}

\def\1h{\hat 1}
\def\2h{{\hat 2}}
\def\3h{{\hat 3}}
\def\4h{{\hat 4}}
\def\be{\begin{eqnarray}}
\def\ee{\end{eqnarray}}
\def\no{\nonumber}

    \def\CF{{\cal F}}

    \def\CK{{\cal K}}

    \def\CN{{\cal N}}
    \def\CO{{\cal O}}

    \def\CT{{\cal T}}

    \def\<{\left\langle\,}
    \def\>{\, \right\rangle}
    \def\[{\left[}
    \def\]{\right]}

    \def\D{{\rm D}}

    \def\sT{{\mathscr T}}

\def\m{\mu}

\def\rrho{{\rho}}

\newcommand{\sbs}[2][r]{\ensuremath{\left(#2\right)_{\mathrm{#1}}}}
\newcommand{\sbsn}[2][r]{\ensuremath{(#2)_{\mathrm{#1}}}}

\renewcommand{\Re}{{\rm Re}\,}
\renewcommand{\Im}{{\rm Im}\,}

\def\i{{\rm i}}

\usepackage{color}

\newcommand{\Wpv}{{\hat {\sW}_{\rm pv}}}
\newcommand{\sW}{{W}}

\newcommand{\hh}{{\hat h}}
\newcommand{\cz}{{\hat *}}

\def\Bm{B_{(-)}}
\def\Rm{R_{(-)}}
\def\Bp{B_{(+)}}
\def\Rp{R_{(+)}}

\DeclareMathOperator{\sign}{sign}
\DeclareMathOperator*{\res}{Res}

\def\sym{$\mathcal N$=4 SYM}

\newcommand{\erase}[1]{{}}

\renewcommand{\url}[1]{\href{#1}{\nolinkurl{#1}}}

\begin{document}

\VerbatimFootnotes

 \begin{frontmatter}
\title{Multiple zeta functions and double wrapping \\ in planar
  $\mathcal{N}=4$ SYM
}

\author[1]{S\'ebastien Leurent}
\ead{s.leurent@imperial.ac.uk}
\author[2,3]{Dmytro Volin}
\address[1]{Theoretical Physics group, Imperial College, South
  Kensington Campus,\\ London SW7 2AZ, United Kingdom}
\address[2]{Nordita,
KTH Royal Institute of Technology and Stockholm University,
Roslagstullsbacken 23
SE-106 91 Stockholm
Sweden}
\address[3]{Bogolyubov Institute for Theoretical Physics,\\ 14-b, Metrolohichna str.
Kiev, 03680, Ukraine}
\ead{vel145@gmail.com}
\begin{abstract}
Using the FiNLIE solution of the AdS/CFT Y-system, we compute 
the anomalous dimension of the Konishi operator in planar \sym\ 
up to eight loops, i.e. up to the leading double wrapping order. At
this order a non-reducible Euler-Zagier sum, $\zeta_{1,2,8}$, appears
for the first time. We find that at all orders in perturbation, every
spectral-dependent quantity of the Y-system is expressed through
multiple Hurwitz zeta functions, hence we provide a {\it Mathematica}
package to manipulate these functions, including the particular case of
Euler-Zagier sums. Furthermore, we conjecture that only Euler-Zagier sums
can appear in the answer for the anomalous dimension at any order in
 perturbation theory.

We also resum the leading transcendentality terms  of the anomalous
dimension at all orders, obtaining a  simple result in terms of Bessel
functions. Finally, we demonstrate that exact Bethe equations should
be related to an absence of poles condition that becomes especially
nontrivial at double wrapping.   
\end{abstract}


\end{frontmatter}


%
%
%
%

%
%

\newpage
\thispagestyle{empty}
\ \\
{\bf Interactive feature:} If you are reading this article as  a pdf
file using a viewer that supports JavaScript (like Adobe Reader), you
can click a sharp symbol, for instance this one 
\PushButton[name=clickD,
 onclick={app.alert('"Whenever you click on a sharp sign, some Mathematica
 code will appear"//Print',3,0,"Greetings, dear reader!")}
]{\#},  whenever you encounter it in the text. When such symbol is
clicked, a pop up window  with a {\it Mathematica} code or example
relevant to the context will appear. The {\it Mathematica} code
requires a number of packages from \cite{Volin:link} \PushButton[name=clickD2$, 
 onclick={app.alert('
In order to load these packages, you can run the following Mathematica code:
PathToFiles = InputString[ "Please enter the name of the folder where you downloaded the packages", $UserDocumentsDirectory];
$Path=DeleteDuplicates[Prepend[$Path,PathToFiles]];
Get /@ {"stylepackage.m", "zetafunctions.m", "FiNLIEKonishi.m", "UpWiseIntegrate.m"};',3,0,"Required
   packages")}
]{\#}.
 More examples
and further explanations how to use packages are given in the notebook \verb#usage.nb# in \cite{Volin:link}.
\tableofcontents
\newpage
\section{Introduction}
During the last decade, there was a remarkable progress in applying
the AdS/CFT integrability to solve the planar $\CN$=4 supersymmetric
Yang-Mills theory (SYM), see \cite{Beisert:2010jr} for a review. In
this context, one of the best studied directions is the AdS/CFT
spectral problem -- the computation of anomalous dimensions of gauge
invariant local operators or, equivalently, of energies of dual string configurations. One typically demonstrates  how to solve
this problem by considering  some  particular operator or a class of
operators; the most known example are the so called twist $J$ spin $S$
states. The shortest  member of this  family whose anomalous dimension $\gamma$
is not protected by supersymmetry is $\Tr ZD^2 Z$, where  $D$ is a
light-cone covariant derivative and $Z$ is a complex scalar field of
\sym. This state known as the Konishi operator  corresponds to $J$=2
and $S$=2. The Konishi operator is  interesting both from the point of view of {\sym} and
its string dual:  on the gauge side it appears among the leading terms of
operator product expansions, and on the string side it is among excitations
with the lowest energy. From the point of view of integrability, 
the anomalous dimension of the Konishi operator is specifically hard
to compute 
perturbatively in the sense that the so called wrapping
corrections \cite{Ambjorn:2005wa} start to contribute earlier than for the anomalous
dimensions of other states. Explicitly, the leading wrapping correction
appears at 
four loops in the perturbative expansion in the 't Hooft coupling
constant $g^2=\frac{g_{YM}^2N_c}{16\pi^2}$. Whereas the anomalous
dimensions of ``long'' operators which are free of wrapping corrections
(e.g. the ones with $J\to\infty$) can be studied by solving the algebraic
equations of the Beisert-Staudacher asymptotic Bethe Ansatz
\cite{Beisert:2005fw,Beisert:2006ez}, the presence of wrapping
corrections requires to solve functional equations instead: the
Gromov-Kazakov-Vieira Y-system \cite{Gromov:2009tv}.

Over the past few years, there has been quick progress in the
understanding of
how to compute anomalous dimensions of ``short'' operators (the ones
that receive wrapping corrections). One already has reasonably precise
numerical values for anomalous 
dimensions of the Konishi state \cite{Gromov:2009zb,Frolov:2010wt}
and of various twist $J$ spin 
$S$ operators  \cite{Frolov:2012zv}. At strong coupling, these
dimensions were found analytically at two \cite{Gromov:2011de} and
three \cite{Gromov:2011bz} loops, with the two-loop result matching
existing  computations \cite{Roiban:2009aa,Roiban:2011fe,Vallilo2011} from the
string theory, whereas at weak coupling the analytical answer for $J=2$ case was
found up to five loops: the results
\cite{Bajnok:2008bm,Bajnok:2009vm,Arutyunov:2010gb,Balog:2010xa} for
$S=J=2$ case coincide
with perturbative quantum field theory computations that reached four
\cite{Fiamberti:2007rj,Fiamberti:2008sh,Velizhanin:2008jd} and five
loop \cite{Eden:2012fe} orders, while the analytical
continuation of the arbitrary $S$ result
\cite{Bajnok:2008qj,Lukowski:2009ce,Balog:2010vf} to $S=-1$  agrees
with the prediction from the  BFKL equation \cite{Kotikov:2007cy}. The
Konishi state was recently analysed even in
greater detail and by now the six \cite{PhysRevLett.109.241601}
and seven loop \cite{Bajnok:2012bz} results for its anomalous
dimension are available. There are also improvements in analysing other operators
than the twist $J$ spin $S$ operators  
\cite{Arutyunov:2011mk,Arutyunov:2012tx},  in studying twisted
\cite{Gromov:2010dy,Arutyunov:2010gu,Bajnok:2010ud,Ahn:2011xq,deLeeuw:2012hp} and $q$-deformed
\cite{Arutyunov:2012zt,Arutyunov:2012ai} versions of the AdS/CFT
spectral problem, and in computing 
angle-dependent cusp anomalous dimensions and related quantities in
{\sym}
from boundary thermodynamic Bethe Ansatz
\cite{Correa:2012hh,Drukker:2012de,Gromov:2012eu}.

Though all these advancements look encouraging, a worrisome sign is
that a number of important results was obtained by approaches that are
stiff for improvement and generalizations.  

At strong coupling, using the method of \cite{Gromov:2011de}  seems
technically infeasible beyond two loops; for instance, the three-loops
result \cite{Gromov:2011bz} was obtained using two-loop findings and
an extra knowledge about the Basso's slope \cite{Basso:2011rs}. This
situation is somehow  similar to the computation of the cusp anomalous
dimension by taking a certain limit of the
generalized scaling function 
\cite{Casteill:2007ct,Gromov:2008en,Volin:2008kd}. While  this approach becomes a
burden beyond two loops, the cusp anomalous dimension  can be computed by other means \cite{Basso:2007wd,Kostov2008a}
to any desired order,
and we might hope to find such means for the dimensions of short
operators as well.

At weak coupling, the analytic five-loop result in
\cite{Balog:2010xa} was obtained by a perturbative
solution of the thermodynamic Bethe Ansatz equations, and there are
conceptual obstacles in generalizing the method of \cite{Balog:2010xa}  to higher loops
\cite{Hegedus:comm}. Another approach of
\cite{Bajnok:2008bm,Bajnok:2009vm} is to use Luscher formulae. It is
an efficient way to account  for the so called single wrapping
effects. However, the Luscher formulae  can be derived only for the
vacuum state, while their generalizations to  excited states (in
particular, to the Konishi state) is a conjecture. At double wrapping
orders, Luscher formulae are known and were used for the vacuum state
of the $\gamma$-deformed theory  \cite{Ahn:2011xq}, however it is not
clear how to proceed and to generalize them to excited states
\cite{Bajnok:comm}. Hence, there is currently a theoretical bound for
applying Luscher formulae, which is seven loops for the case of the
Konishi anomalous dimension and which was reached in
\cite{Bajnok:2012bz}.   

The limitations of these methods and results indicate that some  fundamental
properties of the AdS/CFT integrable system are still not understood
or were not used in these computations. In particular, the above-mentioned
five-loop result at weak coupling  only partially simplifies the original way to
find the spectrum from the thermodynamic Bethe Ansatz equations
\cite{Bombardelli:2009ns,GromovKKV,Arutyunov:2009ur}, whereas, in
later developments, more simpler structures were found behind them: these
equations were shown to be derivable from the Gromov-Kazakov-Vieira
Y-system (or equivalently the  Hirota T-system on
the T-hook), if one supplements it  with certain  constraints
\cite{Cavaglia:2010nm,Balog:2011nm} on discontinuities of the Y-functions.  Then, these
constraints were reduced in \cite{Gromov:2011cx} to  simple
group-theoretical conditions on the  T-functions. 

The T-functions obey the Hirota equation, which is a generalization of
character identities where a dependence on a spectral 
parameter is added. The Hirota equation can be solved by the so called
Wronskian solution, a generalization of the Weyl character formula,
which allows one to parameterize the T-functions in terms of a finite
set of Q-functions. These properties were for a long time known for
various integrable models \cite{Krichever:1996qd}, and their
equivalents were discovered for the case of the AdS/CFT integrability
in \cite{Gromov:2010vb,Gromov:2010km}, see also
\cite{Tsuboi:2009ud,Tsuboi:2011iz,Volin:2010xz} and references therein
for a more generic set up. 

By using these ``algebraic'' findings, the Wronskian parameterization of
the T-functions and the group-theoretical constraints on them, a
finite set of nonlinear integral equations (FiNLIE) which allows one
to compute the anomalous dimensions of several operators was
derived in \cite{Gromov:2011cx} . By contrast,
the thermodynamic Bethe Ansatz equations form an infinite set of
equations. 

In this paper, following our previous work
\cite{PhysRevLett.109.241601}, we apply the FiNLIE, and hence the
 findings discussed above, to show that the weak coupling expansion
can be carried to an arbitrary order in perturbation theory. We
demonstrate this by an explicit computation of the Konishi anomalous
dimension up to eight loops, a benchmark order at which the double
wrapping effects become important for the first time.  

The  goal  of this work was however not only to improve the methods
for weak coupling expansion. Understanding of the AdS/CFT
integrability is far from being ultimate. In particular, it is
conceivable that, instead of the mirror thermodynamic Bethe Ansatz
resulting in an unwieldy infinite set of equations, there should be a
better way to derive the Hirota system and constraints on it. In our
ongoing research \cite{GKLV:wip} we are finding new interesting
features behind the AdS/CFT T-  and Y-systems. On 
the one hand, these features should yield a more efficient way to compute
anomalous dimensions. On the other hand, we hope that they will shed
additional light on the fundamental aspects of the AdS/CFT
integrability which would also  be important beyond the spectral
problem. It appeared that having explicit analytical results is
necessary  to guideline our research, and that not all of the 
effects  manifest themselves  at single wrapping orders. Hence we had
to solve the AdS/CFT Y-system up to a double wrapping order, and we
present such solution in this paper. 

Our  analysis of the perturbative weak coupling behaviour of the FiNLIE
showed that all its quantities can be expressed through the so called
multiple Hurwitz zeta functions. In the following section 
\ref{sec:eta} we acquaint the reader with these functions and with basic
operations on them. Section \ref{sec:weak} describes the FiNLIE
adjusted to weak coupling expansion and gives some simplified examples
which show how to use it. All further technical details are available
online \cite{Volin:link} in the  {\it Mathematica} notebook format. 
In
section \ref{sec:energy} we give our main result: the eight-loop
Konishi anomalous dimension, and in section \ref{sec:tran} we make
some crosschecks by performing an expansion in the inverse powers of
transcendentality of zeta functions. Finally, we summarize our results
in the conclusions section, where we also discuss how the exact Bethe
equations are related to the regularity conditions on a solution of
the FiNLIE. The analyticity properties encoded into these exact Bethe
equations  are an example of a property which acquires new qualitative
features at double wrapping. 

To set up terminology, let us note that we define the anomalous dimension $\gamma$ of an operator as $\gamma=\Delta-\Delta_0$, 
where $\Delta$ is the total conformal dimension and $\Delta_0$ is the classical dimension. For the Konishi operator $\Tr\,ZD^2Z$ on has $\Delta_0=4$. 
The energy $E$ of the  string configuration dual to the Konishi state is defined with respect to the BMN vacuum $\Tr Z^2$ so that $E=\gamma_{\rm Konishi}+2$. The Konishi operator is a member of the so called Konishi supersymmetric  multiplet. Another member of this multiplet is $\sum_{i=1}^6\Tr\,\Phi_i^2$, which is an $R$-symmetry singlet formed from real scalar fields  $\Phi_i$ of \sym\footnote{One defines $Z$ as e.g. $Z=\Phi_1+\i\Phi_2$.}. As all the operators from the Konishi multiplet, this operator has $\gamma=\gamma_{\rm Konishi}$, though its classical dimension is different: $\Delta_0=2$.

\section{\label{sec:eta}Multiple Hurwitz zeta functions}
In section \ref{sec:weak} we will show that all the quantities of the
FiNLIE, and hence of the Y-system, are expressed at any given order of weak
coupling expansion through multiple Hurwitz zeta functions. The goal
of the present section is to define these functions and to list their
essential properties. We start with a specific case of great
importance: Euler-Zagier sums.

\subsection[Euler-Zagier sums]{Euler-Zagier sums\footnote{For a more comprehensive discussion of this subject the reader may consult  \cite{Zudilin,WaldschmidtIntro,Ablinger:2013cf} and references therein.}}
The Euler-Zagier sums, also known as multiple zeta values (MZV), are defined as follows
\begin{equation}
\zeta_{a_1,a_2,\ldots,a_k}=\sum_{0< n_1<n_2<\ldots<n_k<\infty}\frac 1{n_{1}^{a_1}n_{2}^{a_2}\ldots n_k^{a_k}}\,.\label{eq:22}
\end{equation}
$w=\sum\limits_{i=1}^k a_i$ is called the weight, or transcendentality,
of the sum, $k$ is called the depth of the sum. The sum is convergent
for $a_k>1$, $a_k+a_{k-1}>2$, \ldots, $\sum\limits_{i=1}^k a_i>k$. In
the following we will define a regularization which allows us to
define MZV in the marginally divergent case, i.e when strict
inequalities $>$ above are weakened to $\geq$. We  also assume that
$a_i\geq 1$, otherwise the sum can be straightforwardly reduced to
sums of lower depth. For instance: 
\begin{equation}
        \zeta_{-1,a}=\sum_{n_2=1}^\infty\sum_{n_1=1}^{n_2-1}\frac{n_1}{n_2^a}=\frac{1}{2}\left(\zeta_{{a-2}}-\zeta_{a-1}\right)\,.\label{eq:23}
\end{equation}

\paragraph{Stuffle algebra} MZVs form a ring over $\mathbb
Q$. Indeed, consider for example 
\begin{equation}
        \zeta_{a}\,\zeta_b=\sum_{n,m}\frac 1{n^am^b}=\left(\sum_{n<m}+\sum_{n>m}+\sum_{n=m}\right)\frac 1{n^am^b}=\zeta_{a,b}+\zeta_{b,a}+\zeta_{a+b}\,.\label{eq:24}
\end{equation}
Clearly, if we consider two arbitrary MZVs, we can repeat this
logic to split the sum and express their product as a linear combination
of other MZVs. The equalities that we obtain this way are known as
stuffle relations.

\paragraph{Shuffle algebra} At the same time, there is another
inequivalent way to express a product of MZVs through a linear
combination of other MZVs.  For this we use the Feynman representation of
Euler-Zagier sums: 
 \begin{multline}
   \label{FeynmanZeta}
   \zeta_{a_1+1,a_2+1,\ldots,a_k+1}=\\\int\limits_{\infty>t_k>\ldots>t_1>0}\left(\prod_{i=1}^k\frac
     {dt_i}{a_i!(e^{t_i}-1)}\right)(t_1-t_2)^{a_1}\ldots(t_{k-1}-t_k)^{a_{k-1}}t_k^{a_{k}}\,.
 \end{multline}
In the stuffle case we had to split  summation whereas here we will
split domain of integration, e.g\footnote{In this example, both
  $\zeta_1$ (in the l.h.s.) and $\zeta_{2,1}$ are marginally
  divergent. However, the formal manipulations that we write
  disregarding the convergency issue give a correct result if one uses the regularization prescription defined in \appendixref \ref{sec:marg-diverg-eta}. In
  general, the shuffle algebra that we 
  derive this way is valid for a product involving at most one marginally
  divergent MZV in this regularization prescription.}:
\begin{equation}
        \zeta_1\,\zeta_2=\int_{t_1,t_2}d\mu\,t_1^0t_2^1=\int_{t_1<t_2}d\mu\,t_2+\int_{t_1>t_2}d\mu\,\left((t_2-t_1)+t_1\right)=2\zeta_{1,2}+\zeta_{2,1},\label{eq:26}
\end{equation}
where $d\m$ is the measure of integration defined in (\ref{FeynmanZeta}).

Using this logic, we define the so called shuffle relations. This name
comes from the fact that we shuffle the ordering of the integration
variables in all possible ways. In the mathematical literature a
different  integral representation is preferred, the one used to
define multiple polylogarithms \cite{Goncharov}, however the net result is the same.

Both the shuffle and stuffle products are implemented in our {\it
  Mathematica} package \verb#zetafunctions.m# \cite{Volin:link}. \PushButton[name=clickA3,
onclick={app.alert('Example: tmp=zeta[2,2]zeta[5]; {tmp/.subShuffle,tmp/.subStuffle}',3)}
]{\#}

Combinations of shuffle and stuffle products allow one to generate
non-trivial relations between MZVs. A classical example is that the
relations 
\begin{equation}
  \label{eq:28}
  \begin{aligned}
   \zeta_1\zeta_2=&\zeta_{1,2}+\zeta_{2,1}+\zeta_3 &\hspace{2cm}&\textrm{(stuffle product)}\,,\\
  \zeta_1\zeta_2=&2\,\zeta_{1,2}+\zeta_{2,1}&&\textrm{(shuffle product)}
  \end{aligned}
\end{equation}
give the Euler relation
\begin{equation}
       \zeta_{1,2}=\zeta_3\label{eq:29}\,.
\end{equation}

\paragraph{Diophantine conjecture}
\label{sec:dioph-conj}

It is conjectured \cite{Hoffman1997477}  that all
the algebraic
relations between Euler-Zagier sums are generated by shuffle and stuffle relations.
One immediate consequence of this conjecture is that MZVs of different
weight are algebraically independent.  
This conjecture called the \emph{diophantine conjecture} is not
proven. However, one can test it by exploiting integer 
relation algorithms in experimental mathematics (see \cite{BaileyPSLQ}
and references therein). These algorithms allow performing an
efficient and systematic search for  relations of the type $\sum_I
c_I\zeta_I=0$, where the summation is over a set of multi-indices $I$
and $c_I$  are integers. In particular, these algorithms can  show
that no relation exists with coefficients $c_I$-s smaller than a
certain magnitude. The diophantine conjecture found a solid support
using this approach.

Based on the diophantine conjecture, one can show \cite{Hoffman1997477}
that the number $M_w$ of independent irreducible\footnote{MZV is
  called irreducible if it cannot be written  as a combination of MZVs
  of lower depth or weight.}$^,$\footnote{The marginally divergent case is
  not included in this statement, see \appendixref \ref{sec:marg-diverg-eta} for its treatment.}  MZVs
of weight $w$ can be found from the following generating relation
\cite{Broadhurst:1996kc}: $1-x^2-x^3=\prod\limits_{w>0}(1-x^w)^{M_w}$,
which gives the following values of $M_w$ when $w\leq 13$:
\begin{equation}
  \label{eq:12}
  \begin{array}{|c|c|c|c|c|c|c|c|c|c|c|c|c|c|}\hline
w & 1 & 2 & 3 & 4 & 5 & 6 & 7 & 8 & 9 & 10 & 11 & 12 & 13 \\\hline
M_w & 0 & 1 & 1 & 0 & 1 & 0 & 1 & 1 & 1 & 1 & 2 & 2 & 3 \\\hline
  \end{array}\,.
\end{equation}
  
In the eight-loop computation that we present here, it is enough to
find all relations up to transcendentality 13. This is easily done by
brute force based on shuffle and stuffle relations. In
\verb#zetafunctions.m#, the result is saved into the substitution
rules {\verb#subZetaReduce#}
\PushButton[name=clickA4,
onclick={app.alert('Example: zeta[2,4,3]//.subZetaReduce',3)}
]{\#}.  We obtain that
the following MZVs are independent: 
\begin{align}
  \label{eq:32}
  \zeta_2=&\frac{\pi^2}{6},&\zeta_{w}& \textrm{ for }w\textrm{ odd},&
\zeta_{2,6}\,,&&\zeta_{2,8}\,,&&\zeta_{2,10}\,,&&\zeta_{1,2,8}\,&&\zeta_{1,2,10},\,&&\zeta_{1,3,9},\,&&\zeta_{1,1,2,8}\,.
\end{align}
As expected, this is consistent with the values of $M_w$ tabulated in \eqref{eq:12}.
We will actually find out that the weak coupling perturbative
expansion of the Konishi anomalous dimension (\ref{konishianswer}) 
up to eight loops  contains  only  a subset of the  irreducible MZVs
\eqref{eq:32}, namely it only involves $\zeta_{1,2,8}$ and
single-indexed MZVs.

If one wants to go to higher weights, one can use more efficient
algorithms based on  remarkable relations derivable from the shuffle and
stuffle algebras \cite{Ohno199939,HoffmanOhno,CambridgeJournals:414263,Blumlein:2009cf}.
Assuming diophantine conjecture, all the relations among MZVs were worked out in \cite{Blumlein:2009cf} up to weight 22 and, with some restrictions on the depth, to higher weights. These results are available \href{http://www.nikhef.nl/~form/datamine/datamine.html}{online}.

\subsection{Multiple
Hurwitz zeta functions (\texorpdfstring{$\eta$}{eta}-functions)}
\label{sec:eta-functions}
In the evaluation of Feynman integrals, one often encounters
polylogarithms. These functions can be defined as a generalization of
Euler-Zagier sums which preserve the shuffle but not the stuffle algebra, see
e.g. \cite{Goncharov}.  

Quite remarkably, the perturbative weak coupling solution of the
FiNLIE is expressed in terms of functions %
which are in a sense complementary to polylogarithms:
they are generalization of Euler-Zagier sums which preserve the stuffle
but not the shuffle algebra. They are defined as follows 
\begin{equation}
\label{etafunctions}
\eta_{a_1,a_2,\ldots,a_k}(u)=\sum_{0\leq
  n_1<n_2<\ldots<n_k<\infty}\frac 1{(u+{\rm i}\,n_{1})^{a_1}(u+{\rm
    i}\,n_{2})^{a_2}\ldots (u+{\rm i}\,n_k)^{a_k}}\,.
\end{equation}
We introduced ${\rm i}=\sqrt{-1}$ and started the sum from 0 so that
these functions have poles at position $0$, $-{\rm i}$, $-2 {\rm i}$,
$\ldots$. This way, $u$ will coincide with the spectral parameter of
the AdS/CFT integrable model. This sum is  defined under the same restrictions on $a_i$ as for the case of Euler-Zagier sums.

The MZVs are then related to the functions \eqref{etafunctions}
evaluated at the point $u={\rm i}$: 
\begin{equation}
\label{zetaeta}
        \zeta_{a_1,a_2,\ldots,a_k}=\i^{+\sum\limits_{j=1}^ka_j}\eta_{a_1,a_2,\ldots,a_k}(\i)\,.
\end{equation}
The single-indexed functions (\ref{etafunctions})  differ from the Hurwitz zeta
function \linebreak $\zeta_a(u)=\sum_{n>1}(u+a)^{-n}$ only by a slight change of
variable, and they are related to derivatives of the polygamma function $\psi$ by
\begin{equation}
\label{etapolygamma}
 \eta_a(u)=\frac{\i^a}{(a-1)!}\psi^{(a-1)}(-\i\, u)\,.
\end{equation}
Hence, the multiple-indexed functions (\ref{etafunctions}) can be called multiple
Hurwitz zeta functions or multiple polygamma functions.  These sums can be also recast
in terms of infinite cyclotomic harmonic sums defined e.g. in \cite{Ablinger:2011te}. To simplify
further presentation, we will often refer to (\ref{etafunctions})  as $\eta$-functions.

In \appendixref \ref{sec:marg-diverg-eta}, which defines marginally
divergent $\eta$-functions, we choose to enforce \eqref{etapolygamma}
even when $a=1$, which means that we define $ \eta_1(u)\equiv{\i}\psi (-\i\,
u)$. Relation \eqref{zetaeta} is also enforced for marginally divergent case, hence one defines
$\zeta_1\equiv\gamma_{\rm Euler-Mascheroni}$. A number of the FiNLIE functions do depend on this regularization
prescription, however it is a good check of our computation that the
physical quantities, like the anomalous dimension, do not depend on
the regularization.

In the following we will denote multi-indices by the capital letters
$I$,$J$, and $K$. One additional useful property of $\eta$-functions,
which holds in particular for regularized sums is
\begin{align}
        \eta_{a,I}-\eta_{a,I}^{[2]}=&\frac
        1{u^a}\eta_{I}^{[2]}\,,&\textrm{where }\eta_{I}^{[n]}\equiv&
        \eta_{I}%
          (u+{\i} \: n/2%
        ) \,.\label{eq:25}
\end{align}

$\eta$-functions have an infinite ladder of poles at $u=-\i n$,
$n\in\mathbb Z_+$. In the FiNLIE, this structure of poles is natural
at weak coupling $g\ll1$: in this limit, the Zhukovsky
cuts on the intervals $[-2g -\i n,2g -\i n]$ collapse into poles. In
the functions entering the FiNLIE, one can in principle  expect
additional poles at shifted Bethe roots, i.e. at $u=u_j(g)-\i n$. For the
Konishi state, there are two Bethe roots, $u_1(g)=-u_2(g)$, and
$u_1(0)=\frac 1{\sqrt{12}}$. At weak coupling, the appearance of such
poles would indicate the presence of objects like $\eta_I(u+u_j(0))$
or 
even of more complicated functions of type (\ref{eq:9}). However, we
experimentally observe a fine tuning which results in  the
cancellation of such ladders of poles at shifted Bethe roots, even
though a finite number of poles may  survive for some quantities.
Therefore, the possible class of functions entering the FiNLIE solution (at weak
coupling) is  strongly constrained, in the sense that the
poles of $\eta$-functions correspond only to the collapse of Zhukovsky
cuts. Further discussion of this question is given in section
\ref{sec:analytic}.

\subsection{Integrals involving \texorpdfstring{$\eta$}{eta}-functions} 
\label{sec:integration}The integrals we encounter in this work reduce
to a sum of the following elementary ones:
\begin{align}
  \int_{-\infty}^{+\infty}\frac{du}{-2\pi
    \i}\,\bar\eta_{I}^{[-2]}\frac 1{(u-v)^a}\eta_{J}^{[2]}\,,
\label{eq:17}
\end{align}
where $\bar\eta$-functions are the complex conjugate of $\eta$-functions, in the
sense that $\bar\eta_I(u)\equiv\left(\eta_I(u^*)\right)^*$, where $*$
denotes the complex conjugation.

An important property of this integral is that it always evaluates
in terms of $\eta$-functions and MZVs. For example, for $\Im(v)>0$: 
\begin{equation}
\int_{-\infty}^{+\infty}\frac{du}{-2\pi
    \i}\,\bar\eta_{3}^{[-2]}\frac 1{(u-v)}\eta_{2}^{[2]}=-\frac 65\frac{\zeta_2^2}{v}-\i\,\frac{\zeta_3}{v^2}+3\,\i\,\zeta_3\eta_2(v)+\zeta_2\,\eta_3(v)+\eta_{3,2}(v),\label{eq:30}
\end{equation}
and in the limit $v\to 0+\i\, 0$  this integral evaluates to $-\frac{\i}{2}\zeta_5$\,.
\PushButton[name=clickA1,
onclick={app.alert('The result is evaluated from the following input:                                 
rltint=UpIntegrate[-1/(2*Pi*I*(w - (Re[u] + I epsilon)))*etab[3][-2]*eta[2][2],w]/.EndSimplify//.shiftmeta[2]//FuntoMHZbasis;         
{rltint,rltint/.TaylorOptions[Order->0]/.EndSimplify}',3)}
]{\#}

To prove this property, we introduce an intermediate
function that will be called generalized
$\eta$-function\footnote{Generalized $\bar\eta$-functions are defined
  by
  $\bar\eta_{I}^{\{v_1,\ldots\}}(u)\equiv\left(\eta_I^{\{v_1^*,\ldots\}}(u^*)\right)^*$.%
}:
\begin{align}
\eta_{a_1,a_2,\ldots,a_k}^{\{v_1,v_2,\ldots,v_k\}}(u)
\equiv\sum_{0\leq
    n_1<n_2<\ldots<n_k<\infty}\quad
\prod_{j=1}^{n} \left(\frac 1 {u+\mathrm{i}\, n_j - v_j}\right)^{a_j}\,.
\label{eq:9}
\end{align}
The integral \eqref{eq:17} is evaluated iteratively by taking
residues: if the multi-index $I$ has the form $I=b,\check 
I$, and if  $0<\mathrm{Im}(v)<1$
, one has\footnote{For simplicity, we  assume that the integral is
  convergent. In practice, sometimes one needs to combine few
  elementary integrands to achieve convergence. 
}
\begin{align}
  \lefteqn{\int_{-\infty}^{+\infty}\frac{du}{-2\pi
    \i}\,\bar\eta_{I}^{[-2]}\frac 1{(u-v)^a}\eta_{J}^{[2]}}\qquad\nonumber\\=&
  \sum_{k=1}^\infty \int_{-\infty}^{+\infty}\frac{du}{-2\pi
    \i} \left(\frac {\bar \eta_{\check I}^{[-2]}}{u^b}\right)^{[-2 k]}\frac 1{(u-v)^a}\eta_{J}^{[2]}
\no\\=&
\sum_{k=1}^\infty \int_{-\infty}^{+\infty}\frac{du}{-2\pi
    \i} \left(\frac {\eta_{J}^{[2]}}{(u-v)^a}\right)^{[+2 k]} \frac
{\bar \eta_{\check I}^{[-2]}}{u^b} -\res_{u=v}\,\left(\frac {\bar
    \eta_{\check I}^{[-2]}}{u^b}\right)^{[-2 k]}\frac
{\eta_{J}^{[2]}}{(u-v)^a}
\no\\=&\int_{-\infty}^{+\infty}\frac{du}{-2\pi
    \i}\, \frac {\left(\eta_{a,J}^{\{v,0,0,\ldots\}} \right)^{[+2]}
    \bar \eta_{\check I}^{[-2]}}{(u+ i\,0)^b}
-\res_{u=v}\,\bar\eta_{I}^{[-2]}\frac 1{(u-v)^a}\eta_{J}^{[2]} \label{eq:18}\,.
\end{align}
On the one hand, the residue in \eqref{eq:18} is  expressed in terms
of $\eta$-functions at point $u=v$ (because the derivatives of 
$\eta$-functions are themselves $\eta$-functions). On the other hand,
the integral term gives rise to two different cases: if $\check I =
\emptyset$, then by closing the integration contour upwards this
integral evaluates to $-\res_{u=0} \frac {\left(\eta_{a,J}^{\{v,0,0,\ldots\}} \right)^{[+2]}
    \bar \eta_{\check I}^{[-2]}}{u^b}$, whereas if $\check I$ has at
  least one element, it can be written as $\check I=c,\check{\check I}$,
  allowing to move the integration contour again to get
  \begin{multline}
    \label{eq:19}
    \int_{-\infty}^{+\infty}\frac{du}{-2\pi
    \i}\, \frac {\left(\eta_{a,J}^{\{v,0,0,\ldots\}} \right)^{[+2]}
    \bar \eta_{\check I}^{[-2]}}{(u+\i\,0)^b}\\=
    \int_{-\infty}^{+\infty}\frac{du}{-2\pi
    \i}\, \frac {\left(\eta_{b,a,J}^{\{0,v,0,0,\ldots\}} \right)^{[+2]}
    \bar \eta_{\check{\check I}}^{[-2]}}{(u+\i\,0)^c}
-\res_{u=0}\, \frac {\left(\eta_{a,J}^{\{v,0,0,\ldots\}} \right)^{[+2]}
    \bar \eta_{\check I}^{[-2]}}{u^b}\,.
  \end{multline}
If the initial multi-index $I$ has $n$ elements, the  process
stops after $n$ iterations, and one gets the integral
\eqref{eq:17}  in terms of standard $\eta$-functions evaluated either
at point $u=\i$ (where they are equal to MZVs (\ref{zetaeta})) or at
point $u=v$, and of generalized $\eta$-functions of the form 
$\eta_K^{\{\ldots,0,0,v,0,0,\ldots\}}$, 
evaluated at point $u=\i$.

As explained in \appendixref \ref{sec:expr-some-gener}, the generalized
$\eta$-functions of type \linebreak 
$\eta_K^{\{\ldots,0,0,v,0,0,\ldots\}}$ can be expressed through standard
$\eta$-functions and $\bar\eta$-functions evaluated at point
$v$. Moreover, the result of integration is  analytic above or below
the contour of integration and hence cannot contain simultaneously
$\eta$-functions and $\bar\eta$-functions. For instance, in example
\eqref{eq:30} the answer is analytic in the upper half-plane and hence
it should be expressible only through $\eta$-functions. \appendixref
\ref{sec:periodicity} gives an additional 
class of relations which allow one to express $\eta$-functions in terms
of $\bar \eta$-functions (or the opposite) and to insure this last
property we discussed. This class of relations is also necessary 
to show that  terms like the $\eta$-functions evaluated at
Bethe roots cancel in the final expression for the anomalous dimension.

\ \\
The integration algorithm and the properties of $\eta$-functions
described above and in  \appendixref \ref{sec:other-properties-eta}
are implemented in our {\it Mathematica}  
package \cite{Volin:link} which was used to perform
all analytical computations needed 
to solve the FiNLIE at weak coupling. Examples of usage are given
in the file  \verb"usage.nb".

\section{\label{sec:weak}Set of equations for weak coupling expansion}
Our strategy for weak coupling expansion was  summarized in
\cite{PhysRevLett.109.241601}\footnote{ 
The reader might find it useful to take a look on
\cite{PhysRevLett.109.241601} before going into details of this
section}. 
The goal is to determine three
quantities $\rho,\rho_2,U$ from which all the Y-system can be
straightforwardly restored.%
\begin{figure}
  \centering
\includegraphics{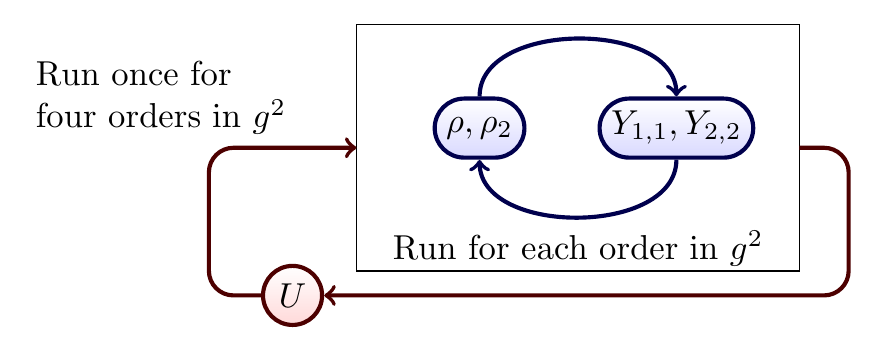}
  \caption{\label{fig:iterative}Structure of the perturbative computation.}
\end{figure}
 As shown in figure \ref{fig:iterative} from
 \cite{PhysRevLett.109.241601} (redrawn above)%
, the set of 
  equations divides into two iterative processes: on the
one hand, there is an
internal  ``{Y-cycle}'' which has to run every time we want to
compute one more order in $g^2$ for the densities 
 $\rho$ and
$\rho_2$. %
On the other hand, there is a ``wrapping
cycle'', which  computes the function $U$, and which only needs
to be run once for every four orders in $g^2$.

Compared to the  weak coupling expansion
\cite{PhysRevLett.109.241601}, which was performed up to six loops, we
will make two essential updates.
 First, we add an equation (\ref{eq:6}) for an auxiliary quantity $\hat h$
 which becomes important at 7 loops. 
   Second, we will compute the wrapping correction to $U$ using
   (\ref{eq:5}) and (\ref{eq:7}), to reach the double-wrapping eight-loop
   order. By 
contrast,  the asymptotic expression for $U$ was sufficient at
single-wrapping.  

For the sake of completeness and to provide some technical details, we
will   write down all the  necessary equations even though some of
them were written in \cite{Gromov:2011cx,PhysRevLett.109.241601}.
Although we will repeat the necessary definitions, the reader can also 
consult \cite{PhysRevLett.109.241601} or 
section 2 of \cite{Gromov:2011cx} for 
some
notations we use, in particular for the definition of integral
kernels, Zhukovsky variables, mirror and magic sheets. The equations
presented below can be used for a  straightforward  weak coupling
expansion to an arbitrary order. Although the expansion is technically
quite evolved, we developed mathematical techniques which allow handling it.

\subsection{\label{sec:wronskian}Restoring the \texorpdfstring{{\rm
      Y}}{Y}- and  \texorpdfstring{{\rm T}}{T}-systems from
  \texorpdfstring{$\rho$, $\rho_2$ and $U$}{rho, rho2 and U}}
The T-(Y-)systems involve infinitely many T-(Y-)functions, and an important
point towards a finite set of equations is to rewrite these functions in terms of a
finite number of Q-functions \cite{Gromov:2010km}, which are
themselves parameterized by a few densities \cite{Gromov:2011cx}. In
the finite set of equations (FiNLIE) of   \cite{Gromov:2011cx}, the T-functions are
parameterized as follows by the functions $\rho,\rho_2,U$:

The T-functions of the right band ($\{T_{a,s}:s\geq a\}$) are computed
in a certain gauge which is denoted by the letter $\cT$. They are
expressed in 
terms of the function $\rho$ alone:
\begin{align}
\label{cT}
  \cT_{0,s}=&1\;, &
\hat \cT_{1,s}=&s+\CK_s\hat\ast \rho\;,&
\hat \cT_{2,s}=&\hat \cT_{1,1}^{[+s]}\,\hat \cT_{1,1}^{[-s]}\;,
\end{align}
where $\hat \cT_{a,s}(u)$ coincides with $\cT_{a,s}(u)$ as soon as
$|\mathrm{Im}(u)|<(1+|s-a|)/2|$, but it has short cuts on $\hat
Z_{\pm s}\equiv [-2g\pm s \frac i 2,2g\pm s \frac i 2]\equiv \hat Z\pm
s \frac i 2$, whereas
$\cT_{a,s}(u)$ has long cuts of the form $\check Z_n\equiv n\frac i 2
+ (-\infty,-2g]\cup[2g,\infty)$. In \eqref{cT}, we use the notations
\begin{align}
(\CK \hat \ast \rho)(u) =&\frac{1}{2\pi
  i}\int_{-2g}^{2g}dv\,{f(v)}/({v-u})\,,& f^{[\pm s]} =&f(u\pm s \frac i 2)
\;,\\
\CK_s \hat \ast \rho=& (\CK
\hat \ast \rho)^{[+s]} - (\CK
\hat \ast \rho)^{[-s]}\,.
\end{align}

On the other hand, the T-functions of the upper band
($\{T_{a,s}~:~a\geq |s|\}$) are 
computed in a different gauge denoted by the letter $\sT$, and they
are parameterized by a handful of 
$q$-functions including $U^2={q_{123}}/{q_1}$:
\begin{equation}
\label{curlyT}
\begin{aligned}
  \sT_{a,2}=&q_{\emptyset}^{[+a]} \bq_\emptyset^{[-a]}\,,\\
\sT_{a,1}=&q_1^{[+a]} \bq_2^{[-a]}+q_2^{[+a]} \bq_1^{[-a]}+q_3^{[+a]} \bq_4^{[-a]}+q_4^{[+a]}
\bq_3^{[-a]}\,,\\
\sT_{a,0}=&q_{12}^{[+a]} \bq_{12}^{[-a]}+q_{34}^{[+a]} \bq_{34}^{[-a]}
-q_{14}^{[+a]} \bq_{14}^{[-a]}-q_{23}^{[+a]} \bq_{23}^{[-a]}-q_{13}^{[+a]}
\bq_{24}^{[-a]}-q_{24}^{[+a]} \bq_{13}^{[-a]}\,,\\
\sT_{a,-1}=&(U^{[+a]}\bar
U^{[-a]})^2\sT_{a,1}\,,\\
 \sT_{a,-2}=&(U^{[a+1]}U^{[a-1]}\bar U^{[-a-1]}\bar
U^{[-a+1]})^2\sT_{a,2}\,.
\end{aligned}
\end{equation}
These $q$-functions are expressed through the functions
$q_{12},q_1,q_2$, and $U$ as follows
\begin{subequations}
\label{eq:3}
\begin{gather}
  \begin{aligned}
\label{eq:qij}
    q_{13}=&q_{12} \sum_{k=0}^{\infty}
    \left(\frac{U^2\,q_1^2}{q_{12}^+q_{12}^-}\right)^{[2k+1]}\,,&
    q_{23}=q_{14}=&q_{12} \sum_{k=0}^{\infty}
    \left(\frac{U^2\,q_1 q_2}{q_{12}^+q_{12}^-}\right)^{[2k+1]}\,,\\
    q_{24}=&q_{12} \sum_{k=0}^{\infty}
    \left(\frac{U^2\,q_2^2}{q_{12}^+q_{12}^-}\right)^{[2k+1]}\,,
  \end{aligned}\\
  \begin{aligned}
    q_{34}=&\frac{q_{13}q_{24}-q_{14}q_{23}}{q_{12}}\,,\hspace{2cm}&
  q_{\emptyset}=&\frac{q_2^--q_2^+}{q_{12}}\,,\label{eq:19b}%
\\
q_3=&\frac{q_2q_{13}^+-q_{14}^+}{q_{12}^+}\,,&
\ q_{4}=&\frac{q_2q_{23}^+-q_{24}^+}{q_{12}^+}\,.
  \end{aligned}
\end{gather}
\end{subequations}
The $\sT$-gauge is chosen such that
\begin{align}
  \label{eq:36}
  q_1=&1\;,&q_{12}=&(u-u_1-\a)(u+u_1+\bar\a)\,,
\end{align}
where $u_1$ is the exact Bethe root and $\a$ is a complex constant,
which is adjusted to enforce the vanishing
$\sT_{1,0}(u_1\pm\i/2)=0$. This constant has order $\a\propto g^8$
an is suppressed by one wrapping (asymptotically, we have
$(q_{12})_{\rm as}=Q\equiv u^2-u_1^2$), hence it can 
be found 
iteratively\footnote{More explicitly, $\sT_{1,0}(u_1+\i/2)=0$ implies
  the following equation:
  \begin{equation}
    \label{eq:37}
    \a=\frac{(\sT_{1,0}(u_1+\frac \i 2)-q_{1,2}^{\vphantom{[0]}} (u_1+\i)\,\bar q_{1,2}^{\vphantom{[0]}}(u_1))+2\,\i\,\,u_1\a^2+\a\,\bar\a\,(-\i+\a+2\,u_1)(\i+\bar\a+2\,u_1)}{-4\,\i\,(u_1-\frac \i 2)\,u_1},
  \end{equation}
where the first term in the numerator is of order $g^8$ and the
remaining ones are of order $g^{16}$. 
}. %
Last, the function $q_2$ is parameterized by
\begin{equation}\label{q2param}
q_2=-\i u+\CK\hat*\rho_2-\CK*\Wpv\;,
\end{equation}
where we denote $(\CK* f)(u)=\frac{1}{2\pi
  i}\int_{-\infty}^{\infty}dv\,{f(v)}/({v-u})$ (resp. $(\CK\hat * f)(u)=\frac{1}{2\pi
  i}\int_{-2 g}^{2 g}dv\,{f(v)}/({v-u})$) and
 where
 we
define $W_{a}\equiv q_{3}^{[+a]}\bar q_{4}^{[-a]}+q_{4}^{[+a]}\bar
q_{3}^{[-a]}$, $\hat\sW\equiv\lim\limits_{a\to0}\sW_a$ for
$u\,\in\check Z$ (this function is then analytically continued to any complex
value of $u$ by avoiding short cuts) and $\Wpv\equiv\frac 12(\hat
W^{[+0]}+\hat W^{[-0]})$ on the real axis
(where $W^{[\pm0]}\equiv \lim_{\epsilon\to 0}W^{[\pm \epsilon]}$)%
. We will also use
$\sW\equiv\lim\limits_{a\to0}\sW_a$ for $u\,\in\hat Z$ later
on\footnote{Note that $\Wpv=\hat W=q_3^{[+0]}\bar
  q_{4}^{[-0]}+q_4^{[+0]}\bar q_{3}^{[-0]}$ for $u\in\check Z$ but
  $\Wpv\neq\,W$ for $u\in\hat Z$.}. The term involving $\sW$ is
suppressed in wrapping and can be found recursively.

The choice (\ref{q2param}) for the parameterization of $q_2$ is
motivated by two features of this parametrization : On the one hand, the
finiteness of the support of $\rho_2$ (i.e. the condition that
$\rho_2=0$ on $\hat Z_0$) is equivalent to the condition
$\sT_{0,1}^c=0$ written in \cite{Gromov:2011cx}, which is a
consequence of $\mathbb Z_4$-symmetry identified in
\cite{Gromov:2011cx}. 
On the other hand, $\rho_2$ is of the form
$\sqrt{4g^2-u^2}f(u)$, where $f$ is an analytic function in
the vicinity of the real axis\footnote{In \cite{Gromov:2011cx}, a slightly different parameterization was used in which $\rho_2$ was  of the
  form $\sqrt{4g^2-u^2}f(u)+(4g^2-u^2)\tilde f(u)$. By contrast, the
  present convention incorporates the  $\tilde f$-term into
   $\CK*\Wpv$.}. The 
latter property is a consequence of the built-in assumption that Q-
and T-functions of the AdS/CFT integrable system have branch points of the second order.

The formulae presented above are nothing but an explicit realization
of the Wronskian solution of the $\mathbb{T}$-hook \cite{Gromov:2010km} bisected
into semi-infinite bands, without taking into consideration any gluing
condition on these bands so far (i.e. that a global gauge  exists in
which the Hirota equation is satisfied everywhere on the
$\mathbb{T}$-hook, including diagonals  $a=\pm s$). The 
parameterization (\ref{cT}) and (\ref{q2param}) was inspired by an
approach in \cite{Gromov:2008gj,Kazakov:2010kf}; it is just a suitable
way to represent the Q-functions respecting basic analytical
properties of the Y-system, with the details about particular state
hidden into  $\rho$ and $\rho_2$. Hence, so far we were exploiting
merely algebraic properties of the Hirota equation, with minor
analytical input. Now we come to more physical constraints which will
determine $U,\rho$, and $\rho_2$ for the Konishi state. 

\subsection{Y-cycle}
We use two integral equations for
$Y_{1,1}$ and $Y_{2,2}$.
First, an %
equation %
for the product  $Y_{1,1}%
Y_{2,2}$ %
is
obtained %
from the analyticity of the ratio
$\frac{1}{ Y_{1,1} Y_{2,2}}\frac{\sT_{1,0}}{\sT_{0,0}^-}$ in the
upper half-plane which  implies   \cite{Gromov:2011cx} the relation\footnote{The real part in the convolution
was denoted
as $\tilde\eta_b$ in \cite{Gromov:2011cx}.} 
\begin{subequations}
\label{eqforYs}
\begin{equation}
\label{eq:1}  \frac{\log\left(\frac{1}{ Y_{1,1} Y_{2,2}}
\frac{\sT_{1,0}}{\sT_{0,0}^-}\right)}{\hat{x}-\frac{1}{\hat{x}}}=\CK
\ast2\Re\left(\frac{\log\left(\frac{\sT_{1,0}}{\sT_{0,0}^+}\right)}{\hat{x}-\frac{1}{\hat{x}}}\right)^{\hspace{-.5em}[-0]}\,,
\end{equation}
where $\hat x=(\sqrt{2+u/g}\sqrt{-2+u/g}+u/g)/2$.
Similarly, from the analyticity of the ratio
$\frac{Y_{1,1}\sT_{0,0}^-}{Y_{2,2}\sT_{1,0}}
\left(\frac{\sT_{2,1}\cT_{1,1}^{-}}{\CT_{1,2}\sT_{1,1}^{-}}\right)^2$
in the upper half-plane, one deduces that the following integral equation for
$\frac{Y_{1,1}}{Y_{2,2}}$ holds:
\begin{multline} \label{eq:2}
  \log\left(\frac{Y_{1,1}}{Y_{2,2}}\frac{\sT_{2,1}^2}{\cT_{1,2}^2}\right)=
\log\left[
  \frac{\sT_{1,0}}{\sT_{0,0}^-}
  \left(\frac {\sT_{1,1}^-} { \cT_{1,1}^- }\right)^2\right]\\
-\ \CK \ast 2\,\i\,\Im\left(\log\left(\frac{\sT_{1,0}}{\sT_{0,0}^+}
\right)-2\log\left(\frac {\sT_{1,1}^+}{ \cT_{1,1}^+ }\right)\right)^{\hspace{-.4em}[-0]}\,.
\end{multline}
\end{subequations}
Introducing
\begin{gather}
  Q=(u-u_1)(u+u_1)\,,\hspace{.7cm}
x=\frac 1 2\left(\frac u g + i\sqrt{4-\frac{u^2}{g^2}}\right)\,,\hspace{.7cm}
\hat x_1^\pm=\hat x(u_1\pm\frac i 2)\,,\nonumber\\
B_{(\pm)}=\prod_{j=1}^2\sqrt{\frac{g}{\hat
    x_j^\mp}}\left(\frac{1}{x}-\hat x_j^\mp\right)
,\hspace{.1cm}
  R_{(\pm)}=\prod_{j=1}^2\sqrt{\frac{g}{\hat x_j^\mp}}\left({x}-\hat
    x_j^\mp\right),\hspace{.1cm}
\hat x_2^\pm=\hat x(-u_1\pm\frac i 2)\,,
\end{gather}
one can show that the asymptotic expressions written in 
\cite{Gromov:2011cx}
\begin{gather}
    \left(Y_{1,1}Y_{2,2}\right)_{\rm as}=\frac{\Bm\Rp}{\Bp\Rm}\,,\quad
\left(\frac{Y_{1,1}}{Y_{2,2}}\frac{\sT_{2,1}^2}{\cT_{1,2}^2}\right)_{\rm
as}=\left(\frac{\Delta_{\rm as}(u_1)}{\Delta_{\rm as}(u_1)-4}\right)^2\frac {Q^+}{Q^-}\frac{\Bp^{[-2]}x^{[-2]}}{\Bm^{[+2]}x^{[+2]}}\,,
\nonumber\\
    (\sT_{a,0})_{\rm as}=Q^{[+a]}Q^{[-a]}\,,\quad(\sT_{a,1})_{\rm as}=a+\CK_s\hat\ast(\rho_2)_{\rm as}\,,\quad(\cT_{1,s})_{\rm as}=s+\CK_s*(\rho)_{\rm as}\,,
\nonumber\\
    (\rho)_{\rm as}=4\,\frac{\sqrt{4g^2-u^2}}{\Delta_{\rm as}(u_1)},\hspace{2cm}
(\rho_2)_{\rm as}=-4\,\frac{\sqrt{4g^2-u^2}}{\Delta_{\rm as}(u_1)-4}\,, 
\label{eq:as1}
\end{gather}
with $\Delta_{\rm as}$ defined in \eqref{wrapen}, solve  (\ref{eq:1})
and (\ref{eq:2})
. Hence we 
subtract them from the exact
quantities in \eqref{eqforYs} and get equations more suitable for weak 
coupling expansion. 

Let us introduce the notation $\sbs F=\frac{F}{\sbs[as]F}$ and define some standard
combinations
\begin{align}
H=&\log \sbs {
\frac{\sT_{1,0}}{\sT_{0,0}^+}},& r=&\log\sbs{\frac{\sT_{1,1}}{\CT_{1,1}}},&
\hat r_*=&\log \sbs{\frac{\hat q_\emptyset}{\hat\CT_{1,1}}}
\label{standardcombinations}
\end{align}
which will be used throughout the text. 
In \eqref{standardcombinations}, $\hat q_\emptyset$ denotes the
function that coincides with $q_\emptyset$ when $\mathrm{Im}(u)>-1/2$ but which
has cuts of the form $\hat Z_{-n}$.

 By subtracting the asymptotic quantities (\ref{eq:as1}) in  (\ref{eqforYs}), one gets
\begin{subequations}
\label{eqforYs2}
\begin{gather}
\log\sbs{Y_{1,1}
    Y_{2,2}}=\bar{H}-\left(\hat{x}-\frac{1}{\hat{x}}\right)
\CK\ast
\left(\frac{\bar{H}^{[+0]}}{\hat{x}^{[+0]}-\frac{1}{\hat{x}^{[+0]}}}+
\frac{H ^{[-0]}}{\hat{x}^{[-0]}-\frac{1}{\hat{x}^{[-0]}}}\right)
,\\
\log\sbs{\frac{Y_{1,1}}{Y_{2,2}}\frac{\sT_{2,1}^2}{\cT_{1,2}^2}}=
\bar H+2r^-+\ \CK \ast\left(H^{[-0]}-\bar H^{[+0]}+2r^{[1-0]}-2r^{[-1+0]}\right)\,.
\end{gather}
\end{subequations}
The equations (\ref{eqforYs}) and (\ref{eqforYs2}) are valid in the
upper half-plane. The latter ones are used in  derivation of
equation (\ref{eq:5})  on $(U)_r$. Together with their complex conjugate,
they also determine $(Y_{2,2}^+/Y_{2,2}^-)_r$ which appears in the
exact Bethe equations (\ref{eq:BetheAn}).

To find
the densities $\rho$ and $\rho_2$, one  needs in particular to
analytically continue  
(\ref{eqforYs2})   to the domain where $u$ is real and belongs to $[-2
g, 2g]$. In that case, the analytical continuation picks a residue
which cancels some inhomogeneous terms in the equations and one
gets\footnote{We  use the parity
  $\sT_{a,s}(-u)=\sT_{a,s}(u)\,,\cT_{a,s}(-u)=\cT_{a,s}(u)\,,$ 
which is satisfied for the Konishi state, and the reality of
T-functions in the chosen gauges. These properties imply that
$H(-u)=\bar H(u)$ and $r(u)=\bar r(u)=r(-u)$.}: 
\begin{subequations}
\label{dsYs}
\begin{align}
\label{dsYsa}
&\log\sbs[r]{{Y_{1,1}Y_{2,2}}}=-4\,\i\,g\,\sqrt{1-z^2}\int _{-\infty -i 0}^{+\infty -i 0}\frac{\mathrm{d}v}{-2\pi\, i}\frac{1}{(2\,g\,z)^2-v^2}\frac{H}{\sqrt{1-\frac{4g^2}{v^2}}}, \\
\label{dsYsb}
&\log \sbs[r]{\frac{Y_{1,1}}{Y_{2,2}}\frac{\sT_{2,1}^2}{\cT_{1,2}^2}}=
\int _{-\infty -i 0}^{+\infty -i 0}\frac{\mathrm{d}v}{-2\pi
  i} \frac{2v}{(2\,g\,z)^2-v^2} \left(H+ 2\, r^+\right)\,,
\end{align}
\end{subequations}
where we use $z=u/2\,g$, and the weak coupling expansion is to be
performed in the double scaling regime: $g,u\to 0$ with $z$ being
finite. In  \eqref{dsYs}, $H$ (resp. $r^+$) implicitly denotes $H(v)$ 
(resp. $r(v+\i/2)$), and starting from now, for all functions that
appear inside an integral, if the argument of a function is not
specified then it is the integration variable.

Equations (\ref{eqforYs}) on $Y_{1,1},Y_{2,2}$ correspond to the
physical constraints, and in particular one can
show their equivalence to the corresponding TBA equations
\cite{Gromov:2011cx}. On the other hand, there is a purely algebraic
constraint which tells us that $Y_{1,1},Y_{2,2}$ can be expressed through
$\rho,\rho_2$, and $U$ (where $U$ is suppressed in
wrapping and in this sense is  known). Explicitly
\begin{subequations}
\label{magic}
\begin{eqnarray}
\label{magicup}
\frac{1+Y_{2,2}}{1+Y_{1,1}^{-1}}&=&%
{\frac{\sT_{2,2}^+\sT_{2,2}^-}{\sT_{3,2}\sT_{1,2}}}\,/\,{\frac{\sT_{1,1}^+\sT_{1,1}^-}{\sT_{1,0}\sT_{1,2}}}=\frac{q_{\emptyset}^{+}\bar
  q_{\emptyset}^-\sT_{1,0}}{\sT_{1,1}^+\sT_{1,1}^-},\\
\frac{1+Y_{2,2}^{-1}}{1+Y_{1,1}}&=&%
{\frac{\cT_{2,2}^+\cT_{2,2}^-}{\cT_{2,1}\cT_{2,3}}}\,/\,{\frac{\cT_{1,1}^+\cT_{1,1}^-}{\cT_{0,1}\cT_{2,1}}}=\frac{\hat\cT_{1,1}^{[1+0]}\hat\cT_{1,1}^{[-1-0]}}{\cT_{1,1}^+\cT_{1,1}^-}\;.
\label{magicright}
\end{eqnarray}
\end{subequations}
One can check that these equations are also satisfied by the
asymptotic solution (\ref{eq:as1}), so we also subtract   the
asymptotic quantities to  prepare for a weak coupling
expansion. Finally, by excluding $Y_{1,1}$ and $Y_{2,2}$ from
(\ref{magic}) and (\ref{dsYs}) and performing some 
algebra one gets 
\begin{subequations}\label{eq:39}
\begin{multline}
\hat r_*^{[1+0]}+\hat{\bar{r}}_*^{[-1-0]}-r^+-r^-+\log\sbs[r]{\sT_{1,0}}=
\\-4\,\i\,g\,\sqrt{1-z^2}\int _{-\infty -i
0}^{+\infty -i 0}\frac{\mathrm{d}v}{-2\pi\, i}\frac{1}{(2\,g\,z)^2-v^2}\frac{H}{\sqrt{1-\frac{4g^2}{v^2}}},
\end{multline}
\begin{multline}
\hat r_*^{[1+0]}+\hat{\bar{r}}_*^{[-1-0]}+r^++r^-+\log\sbs[r]{\sT_{1,0}}+2\log\sbs[r]{\frac{\rho}{\rho_2+W-\Wpv}}=
\\+\int _{-\infty -i 0}^{+\infty -i 0}\frac{\mathrm{d}v}{-2\pi
  i} \frac{2v}{(2\,g\,z)^2-v^2} \left(H + 2\, r^+\right)\,.
\end{multline}
\end{subequations}
This is a set of two coupled equations which constitute the
Y-cycle. They are used to find $\rho$ and $\rho_2$. To uniquely solve
them, an additional constraint $\sT_{1,1}(u_1)=0$ should be
imposed. An explicit 
expansion of these equations is worked out in the {\it Mathematica}
notebook \verb#WeakCoupling.nb# \cite{Volin:link}, while here we will
discuss an instructive simplified example, by putting
$W_a=U=\a=H=0$. We have put to zero the quantities that are of order
$g^8$, i.e. the quantities which are responsible for the  wrapping
corrections. In this approximation we have $\hat r_*=\hat r$, and the
equations \eqref{eq:39}  reduce to 
\begin{subequations}
\label{withoutwrap}
\begin{flalign}
&     \hat r^{[1+0]}+\hat{{r}}^{[-1-0]}-\hat
     r^{[1-0]}-\hat{{r}}^{[-1+0]} =0\,,\\
&\hat r^{[1+0]}+\hat{{r}}^{[-1-0]}+\hat r^{[1-0]}+\hat{{r}}^{[-1+0]}
\no\\
&\hspace{7em}+\,2\log\sbs[r]{\frac{\rho}{\rho_2}}-\int _{-\infty -i 0}^{+\infty -i 0}\frac{\mathrm{d}v}{-2\pi
  i} \frac{4\,v\, \,\hat r^+}{(2\,g\,z)^2-v^2}=0 ,\label{withoutwrapb}
\end{flalign}
\end{subequations}
with
\begin{align}
\hat r=&\log\sbs[r]{\frac{\hat q_2^{[+1]}-\hat q_{2}^{[-1]}}{1+\hat c^{[+1]}-\hat c^{[-1]}}}\,,
\end{align}
where $\hat q_2=-\i\,u+\CK\hat\ast\rho_2$, $(\hat q_2)_{\rm
  as}=-\i\,u+\i\frac{c_1}{g\,\hat x}$, $\hat c=\CK\hat*\rho$ and
$\sbs[as]{\hat c}=\i\,\frac{g\, c_2}{\hat x}$ are defined everywhere
in the $u$-plane outside the cut $\hat Z_0$. The constants $c_i$ are
given by $c_1=-\frac{4g^2}{\Delta_{\rm as}-4},
c_2=\frac{4}{\Delta_{\rm as}}$, but we will only use that
$c_i\propto\CO(g^0)$.

Next, we define perturbations $\delta \hat q_2=\hat q_2-\sbs[as]{\hat q_2}$, $\delta
\hat c=\hat c-(\hat c)_{\rm as}$,
$\delta\rho_2=\rho_2-\sbs[as]{\rho_2}$, $\delta\rho=\rho-\sbs[as]\rho$
and expand the equations to the 
linear order in the perturbations and to the leading
nontrivial order in $g$. To perform the expansion, one uses firstly
that close to the cut one has  
\begin{align}
     (\hat q_2)_{\rm as}^{[\pm 0]}=&\i\frac{c_1}{g\,\hat x^{[\pm 0]}}+\CO(g^1)\,,& (\hat c)_{\rm as}^{[\pm 0]}=&\i\frac{g\, c_2}{\,\hat x^{[\pm 0]}}+\CO(g^3)\,,
\end{align}
with $\hat x^{[\pm 0]}=1/\hat x^{[\mp 0]}=z\pm\i\sqrt{1-z^2}$\,, and
\begin{equation}\label{eq:hatcpm}
\delta\hat c^{[\pm 0]}(z)=\pm\frac 12\delta\rho(z)+\pint_{-1}^1\frac{d\tilde z}{-2\pi\i}\frac{\delta\rho(\tilde z)}{z-\tilde z}\,,
\end{equation}
where we view $\delta \rho$ as a function of $z=u/2g\in[-1,1]$.
One will find that only a certain combination of $\delta\hat c$,
namely $\delta\hat c^{[+0]}-\delta\hat c^{[-0]}=\delta\rho$, appears
at the leading order. For $\delta\hat q_2^{[\pm 0]}$ one can use the
representation which is analogous to (\ref{eq:hatcpm}), however it is
more instructive to keep $\delta\hat q_2^{[\pm 0]}$ as it is. 

Secondly, one  has the following perturbative expansion if the
argument of a function is far from the cut: 
\begin{align}\label{ordscexp}
(\hat q_2)_{\rm as}(u)=&-\i\, u+\i\,\frac{c_1}{u}+\CO(g^1),\ \ (\hat c)_{\rm as}=\CO(g^2),\no\\
\delta\hat q_2(u)=&\frac{2g}{-2\,i}\int_{-1}^{1}\frac{dw}{\pi}\frac {\delta\rho_2(w)}{u-2 gw}=\frac{\i\, g}{u}\,M[\delta\rho_2]+\ldots\,,
\end{align}
where we defined $M[f]\equiv\int_{-1}^1\frac{dz}{\pi}f(z)$.

By comparing \eqref{eq:hatcpm} to \eqref{ordscexp}, we see that, far
from the cut, these functions are suppressed by one order of $g$
compared to the same functions close to the cut. In particular,
$\delta\hat c$ with its argument far from the cut does not appear in
the leading order expansion. 

The argument of a function is far from the cut in two situations. The
first one is  when one encounters a function with a shift: $f^{[\pm
  2]}=f(2gz\pm\i)=f(\pm\i)+\ldots$\ . The second one is when we
integrate over the argument, as in the integral in
(\ref{withoutwrapb}). The contour of integration can be  deformed such
that it never approaches the branch points $u=\pm 2g$. Hence one can
apply (\ref{ordscexp}) and take the integral in (\ref{withoutwrapb})
explicitly.

By implementing the outlined strategy, one gets  at the leading order
\begin{align}
    (\hat x^{[+0]})^2\,\delta \hat q_2^{[+0]}-(\hat x^{[-0]})^2\,\delta\hat
q_2^{[-0]}=\lambda_1\delta\rho
=\i\,\lambda_2\, M[\delta\rho_2](\hat x^{[+0]}-\hat x^{[-0]})\,,
\label{finalrhoeqns}
\end{align}
where $\lambda_1=\frac{c_1(2\,c_1\,c_2-4\,c_1-1)}{c_2(1-2c_1)}$ and $\lambda_2=\frac{c_1(2\,c_1\,c_2-4\,c_1-1)}{(1+4c_1)(c_1c_2-c_1-1)}$.

Note that the equation $(\hat x^{[+0]})^2\,\delta \hat q_2^{[+0]}-(\hat x^{[-0]})^2\,\delta\hat
q_2^{[-0]}=0$ always has a solution $\delta{\hat q}_2=\frac{A}{\hat
  x^2}(\hat x+\frac 1{\hat x})$ which has good analytic and parity
properties (it decreases at infinity and it is antisymmetric on the
magic sheet). Hence, for any constant $\tilde M$ one can solve
(\ref{finalrhoeqns}) by
\begin{equation}
\delta \hat q_2=\frac {\i\, \lambda_2\tilde M}{\hat x}+\frac{A}{\hat x^2}\left(\hat x+\frac
1{\hat x}\right)\,,
\end{equation}
where $A$ is adjusted such that $M[\delta\rho_2]=\tilde M$. This
one-parametric ambiguity is fixed by recalling that one should impose
$\sT_{1,1}(u_1)=0$. Indeed, since $\sbs[as]{\sT_{1,1}}$ vanishes at
Bethe roots by construction, one should impose that the correction
to $\sbs[as]{\sT_{1,1}}$ obeys:
\begin{equation}\label{eq:10101}
\delta\sT_{1,1}(u_1)=\left.\delta q_2^++\delta\bar
  q_2^-+W_1\right|_{u=u_1}=\frac{g}{u_1^2+\frac 14}
M[\delta\rho_2]+\ldots \, =0,
\end{equation}
where $\ldots$ denotes corrections of higher order in $g$. These
corrections  define the value of $M[\delta\rho_2]$, and in particular
the magnitude of $\delta\rho_2$. In our simplified case $W_1=0$ and
hence $M[\delta\rho_2]=0$, which means, due to (\ref{finalrhoeqns}),
that $\delta\rho_2=\delta\rho=0$. This is only expected because the asymptotic solution should be exact when we set to
zero all the terms that are responsible for wrapping. Although the
solution we've got is trivial, its derivation allowed us to demonstrate
the essential features of the perturbative expansion of (\ref{eq:39}). 

When we add the wrapping corrections, they produce non-zero terms in the
r.h.s. of (\ref{withoutwrap}) which are suppressed in powers of $g$
and induce corrections to the asymptotic expressions for $\rho$
and $\rho_2$. At the leading order one has
\begin{equation}
  \label{eq:40}
\begin{aligned}
\delta\rho=&g^9\sqrt{1-z^2}\,(-162-432\,\zeta_3+432\,\zeta_5+504\,\zeta_7),\\
\delta\rho_2=&g^7\sqrt{1-z^2}\,\left[\vphantom{\left(\frac 5 7\right)}
-378-864\zeta_3+\frac{72}{5}\zeta_2^2+48\zeta_3^2+\frac{656}{35}\zeta_2^3+624\zeta_5 +168\zeta_7  \right.\\
&\qquad +\Im (-144 \sqrt{3} \eta0_2^++144 \eta0_3^+-24 \sqrt{3} \eta0_4^+)
+z^2\left(432-432\zeta_2
\vphantom{\frac 5 7}\right.\\
&\qquad\quad\left.\left.+144\zeta_3+\frac{864}{5}\zeta_2^2+576\zeta_3^2+\frac{7872}{35}\zeta_2^3-1536\zeta_5-672\zeta_7\right)\right],
\end{aligned}
\end{equation}
where $\eta0_a^+\equiv\eta_a(\frac 1{\sqrt{12}}+\frac{\i}2) $.

Since $\frac{\rho(u)}{\sqrt{4g^2-u^2}}$
and
$\frac{\rho_2(u)}{\sqrt{4g^2-u^2}}$ are even functions analytic in the
vicinity of the real axis, they are represented in the double scaling
regime as finite degree polynomials in $z^2$ at any given order of
$g$, c.f. \eqref{eq:40}. This property eventually follows from (\ref{finalrhoeqns})
supplemented with sources from wrapping terms, and it can be
used to transform (\ref{finalrhoeqns}) to  algebraic equations for the
coefficients of the polynomials.

 In  \verb+WeakCoupling.nb+ \cite{Volin:link}, we  compute the
first three orders of $\delta\rho$ and $\delta\rho_2$, $\delta \rho$ up to the order $g^{13}$
and $\delta \rho_2$ up
to the order $g^{11}$. These orders are necessary for our computation
of the energy up to eight loops. One formally  needs also the fourth order of $\delta\rho_2$, however it enters only as $M[\delta\rho_2]$, because of expansion similar to (\ref{ordscexp}), and hence it is found from (\ref{eq:10101}).

\subsection{Wrapping cycle}

An  equation on the function
$U$, which constitutes the wrapping cycle, is  \cite{Gromov:2011cx}:
  \begin{gather}\label{eq:4}
    \left[\frac{U}{\hh} \frac{\hh^{[2]}}{ U
        ^{[2]}}\right]^2=\left(\frac{Y_{1,1}\sT_{0,0}^-}{Y_{2,2}\sT_{1,0}}
    \[\frac{\sT_{2,1}\cT_{1,1}^{-}}{\CT_{1,2}\sT_{1,1}^{-}}\]^2\right)\times\left( \frac{{Y_{1,1}Y_{2,2}}\sT_{0,0}^-}{\sT_{1,0}}
    \right)^{[2]}\,,
    \end{gather}
where on the r.h.s we outlined two factors encountered
previously in the text. Both of them are analytic in the upper
half-plane, where (\ref{eq:4}) is defined. 

This equation on $U$ contains an auxiliary function $\hat h$ which
also appears in the Bethe equations and which is found from: 
\begin{align}
&\log \hh=-2\log{\hat x}+ {\cal
      Z}\cz\log\left(\frac{\CF^{+}(Y_{1,1}Y_{2,2}-1)}{\rrho}\right)\,, \label{eq:6}
\\
&\label{eqonF}\CF^+=\Lambda _F\prod _j\left(i \cosh\left[\pi
    \left(u-u_j\right)\right]\right)\exp\left[\vphantom{\int _{\mathbb{R}\setminus[-2g,2g]}}\right.
\\
&\left.\hspace{5em}\int _{\mathbb{R}\setminus[-2g,2g]}\frac{\mathrm{d}v}{2i}(\coth[\pi (u-v)]+\sign[v])\log\left[Y_{1,1}Y_{2,2}\right]\right]\,.
\no
\end{align}
Equation (\ref{eq:4}) actually allows finding $U$ only up to a normalisation which is fixed
by 
the following additional condition: %
\begin{align}
U\bar U=&\sqrt{\sT_{0,0}^+\sT_{0,0}^-}\,\,\frac{1-Y_{1,1}Y_{2,2}}{\rho_2+W-\Wpv}\,,
&u\in \hat Z\,.\label{eq:7}
\end{align}
For this reason the normalization constant $\Lambda_F$ is inessential.

The asymptotic expressions for $\hat h$ and $U$ are given by
\begin{align}\label{as:2}
     \sbsn[as]{\hat h}=&\hat x^{-2}\hat \sigma_1,&\sbsn[as]U=&\Lambda_U\frac{B_{(-)}}{x}\left(\frac{B_{(-)}}{B_{(+)}}\right)^{\frac {\D^2}{1-\D^2}}\sigma_1,
\end{align}
where $\frac{\hat \sigma_1(u+\i/2)}{\hat
  \sigma_1(u-\i/2)}=\prod\limits_{j=1}^2\sigma_{\rm BES}(u,u_j)$ is
the BES dressing phase \cite{Beisert:2006ez}. 

Let us sketch how these asymptotic expressions are
derived. To this end, let us note that equation (\ref{eq:6}) is
equivalent to the Riemann-Hilbert problem $\hat 
h^{[+0]}\hat h^{[-0]}=\frac{\CF^+(1-Y_{1,1}Y_{2,2})}{\rho}$, under the
boundary condition $\hat h \sim u^{-1-L/2}$ at $u\to\infty$. By
considering $\hat h$ as function of $\hat x$ (which is justified
because $\hat h$ has only one cut on the physical sheet), and using
the periodicity of $\cF$, one can derive from this Riemann-Hilbert problem
the equation $\frac{\hat h(\hat x^{+})\hat h(\frac 1{\hat
    x^+})}{{\hat h(\hat x^{-})\hat h(\frac 1{\hat
      x^-})}}=\frac{\rho^-}{\rho^+} \frac{1-Y_{1,1}^+Y_{2,2}^+}
{1-Y_{1,1}^-Y_{2,2}^-}$. 
For the asymptotic values (\ref{eq:as1}) of $\rho$ and Y-s, this is
precisely the crossing equation \cite{Janik:2006dc} on the dressing
phase  analytically continued using the trick of
\cite{Volin:2009uv}. This explains how the expression (\ref{as:2}) for
$\sbsn[as]{\hat h}$ is obtained. To simplify (\ref{eq:4}) for asymptotic
quantities and  to find $(U)_{\rm as}$, one uses the following
remarkable relation: $\left(\frac{\sT_{1,1}}{\cT_{1,1}}\right)_{\rm
  as}=\hat x_1^+\hat x_1^-\frac{B_{(+)}^-x^-}{B_{(-)}^+x^+}$\,.

Like for the {Y-cycle}, we  subtract the asymptotic solution
from \eqref{eq:4} to get 
\begin{multline}
  \label{eq:5}
  \log\sbs{\frac{{\hh}^{[2]}}{{\hh}
      }\frac{U}{U^{[2]}}}=
\int_{-\infty-\i0}^{+\infty-\i0}\left[
\vphantom{\frac{\left(-\frac{u+i}{g}\sqrt{1-\frac{4g^2}{(u+i)^2}}\right)}{\frac{v}{g}\sqrt{1-\frac{4g^2}{v^2}}}}
\left(\frac{1}{u-v}-\frac{1}{v+u}\right) \left(\frac {H} 2 +
  r^+\right)
\right.\\ +
\left.
\left(\frac{1}{u-v+i}-\frac{1}{u+v+i}\right)\frac{\left(-\frac{u+i}{g}\sqrt{1-\frac{4g^2}{(u+i)^2}}\right)}{\frac{v}{g}\sqrt{1-\frac{4g^2}{v^2}}}\frac{H}2\right]\frac{\mathrm{d}v}{-2 \i \pi}\,.
\end{multline}
\begin{multline}\label{eq:hh}
\log(\hat h)_r= {\cal
      Z}\cz\left[\log\left(\frac{Y_{1,1}Y_{2,2}-1}{\rrho}\right)_{\rm r}-\frac 12\log(Y_{1,1}Y_{2,2})_{\rm r}\vphantom{\int\frac{dv}{2\i}}\right.\\
\left.+\int\frac{dv}{2\i}\left((\coth(\pi(2gw-v))-\coth(\pi\,v))\log(Y_{1,1}Y_{2,2})_{\rm r}\right)_{\rm pv}\vphantom{\log\left(\frac{Y_{1,1}Y_{2,2}-1}{\rrho}\right)_{\rm r}-\frac
12\log(Y_{1,1}Y_{2,2})_{\rm r}}\right],
\end{multline}
where the expression $\log(Y_{1,1}Y_{2,2})_r$ should be computed in
the double scaling regime $u=2gz$, using (\ref{dsYsa}).
Note that
\begin{equation}
\label{appr1}
\log(Y_{1,1}Y_{2,2}-1)_r=\sbs[as]{\frac{Y_{1,1}Y_{2,2}}{Y_{1,1}Y_{2,2}-1}}\log(Y_{1,1}Y_{2,2})_r+\ldots,
\end{equation}
and the subleading terms of this expansion (denoted by $\ldots$) are
not necessary for the eight-loop computation in this paper.

Finally, the equation for the normalization of $U$ (\ref{eq:7}) should
be transformed into an equation for the normalization of $(U)_r$,
which requires to continue (\ref{eq:5}) to real values of $u$ and to
re-expand it in the double scaling regime $u=2\,g\,z$. For the leading
wrapping correction of $U$, this gives the following equation for the
normalization of $U$: 
\begin{multline}\label{eq:11}
\log(U^{[2]}\bar U^{[-2]})_{\rm r}=\log\sbs[r]{\frac{1-Y_{1,1}Y_{2,2}}{\rho_2+W-\Wpv}\sT_{1,0}}+r^++r^-
\\+2\int_{-\infty-\i\, 0}^{+\infty-\i\, 0}\frac{dv}{-2\pi\i}\left(\frac{H+2r^+}{v}-\frac{iH}{1+v^2}\right)\,.
\end{multline}
The dependence on $z$ should be the same in the r.h.s and the l.h.s., %
which is used as a nontrivial check for our computations.
 
Let us note that $\log\sbsn\hh=\mathrm{constant}\cdot g^8 +
\mathcal{O}\left(g^{10}\right)$. Since an overall normalization of $\hat h$
is irrelevant, \ $\log\sbsn\hh$ does not contribute to the 
computation of $U$ at eight loops. However, the terms in $\log\sbsn{\hat h}$ of order
$g^{10}$ and $g^{12}$  are needed to find
corrections to the Bethe equations, as we will see below.

\subsection{Bethe equations} The Bethe equations can be written as
\begin{align}
\label{eq:BetheAn}
  e^{\phi(u_j)}=&1\,,&\textrm{where }
e^{\phi(u)}\equiv& -\left(\frac{\hat h^-}{\hat  h^+}\right)^2
\frac{Y_{2,2}^+}{Y_{2,2}^-} \frac{\CT_{1,2}^+}{\CT_{1,2}^-} \frac{\hat\CT_{1,1}^{[-2]}}{\hat\CT_{1,1}^{[+2]}}\,.
\end{align}
The asymptotic function $\sbsn[as]\phi$ precisely coincides with
the logarithm of the Beisert-Staudacher asymptotic Bethe
equations \cite{Beisert:2005fw,Beisert:2006ez}. The function $\phi$ is
subject to corrections to its asymptotic value; these corrections are 
computed from the following equation:
\begin{equation}\label{exBetherat}
{\delta\phi}=\log\left(\left(\frac{\hat h^-}{\hat h^+}\right)^2\frac{\sqrt{ \frac{Y_{1,1}^+Y_{2,2}^+}{Y_{1,1}^-Y_{2,2}^-}
}}{\sqrt\frac{\left(\frac{Y_{1,1}}{Y_{2,2}}\frac{\sT_{2,1}^2}{\CT_{1,2}^2}\right)^+}{\left(\frac{Y_{1,1}}{Y_{2,2}}\frac{\sT_{2,1}^2}{\CT_{1,2}^2}\right)^-}}\frac{\sT_{2,1}^+}{\sT_{2,1}^-}\frac{\hat\cT_{1,1}^{[-2]}}{\hat\cT_{1,1}^{[+2]}}\right)_{\rm
r}\,.
\end{equation} 
The square root in the numerator (resp. the denominator) is expressed
from (\ref{dsYsa}) (resp. from (\ref{dsYsb})), and $\hat h$ is
expressed from (\ref{eq:hh}), while $\hat \cT_{1,1}$ and $\sT_{2,1}=q_2^{[2]}+\bar
q_2^{[-2]}+W_2$ are found from the Wronskian parameterization
discussed in section \ref{sec:wronskian}. 

What one really needs is the value of $\delta\phi$ at $u=u_1$ which we
parameterize as follows: 
\begin{equation}
        \delta\phi(u_1)=g^{8} m_1+g^{10}m_2+g^{12}m_3+g^{14}m_4
+\ldots\label{eq:10}\,.
\end{equation}
We explicitly computed $m_1$,$\ldots$, $m_4$ which are used for
our computation of 
the energy up to eight loops. Their expressions can be found in \verb$WeakCoupling.nb$ \cite{Volin:link}. Using this data, one finds the
exact position of the Bethe root if one notices that, since
$\phi(u_1)=0$, one has
$\delta\phi(u_1)=\phi(u_1)-\sbsn[as]\phi(u_1)=-\sbsn[as]\phi(\tilde
u_1+\delta u_1)$. The latter equation is solved perturbatively as
follows 
\begin{multline}
        u_1=\tilde u_1+\i\left(-\frac{1}{9}\, g^8-\frac{4}{9}g^{10}+2g^{12}-8g^{14}\right)(m_1+g^2m_2+g^4m_3+g^6 m_4)\\+\CO(g^{16})\,,
\end{multline}
      where $\tilde u_1$ is the solution of the asymptotic Bethe
      equation $\sbsn[as]\phi(\tilde u_1)=0$. Its explicit expression
      is 
\begin{multline}\begin{aligned}
\tilde u_1=\frac 1{\sqrt{3}}\left[\,\vphantom{\frac
    12}\right.&
\frac 12+4\,g^{2}-10\,g^{4}+8\,g^{6}(7+3\,\zeta_{3})+\,g^{8}(-461-240\zeta_{5}-144\zeta_{3})
\\
&+4\,g^{10}(1133+252\,\zeta_{3}+378\,\zeta_{5}+630\,\zeta_{7})
\\
&-6\,g^{12}(7945+1556\,\zeta_{3}+48\,\zeta_{3}^{2}+1944\,\zeta_{5}+2772\,\zeta_{7}+4704\,\zeta_{9})
\\
&+24\,g^{14}(21577+4572\,\zeta_{3}+240\,\zeta_{3}\,\zeta_{5}+24\,\zeta_{3}^{2}+4784\,\zeta_{5}
\end{aligned}
\\
\left.+5706\,\zeta_{7}+8064\,\zeta_{9}+13860\,\zeta_{11})\vphantom{\frac
    12}+\ldots\right]\,.
\end{multline}

\subsection{Summary of the perturbative expansion} %
To conclude, the analytic perturbative
solution of the FiNLIE relies on a perturbative expansion of the Bethe
root $u_1$, the two densities $\rho$ and $\rho_2$, and the function
$U$. 

In practice, one computes the deviations from the asymptotic
expressions: $\delta\rho=\rho-\sbs[as]\rho$,
$\delta\rho_2=\rho_2-\sbs[as]{\rho_2}$, using equations (\ref{eq:39}),
and   
$\log\sbs[r]U=\log U/\sbs[as]U$, using equations (\ref{eq:5}) and
(\ref{eq:7}). These equations contain a number of auxiliary
objects. Firstly,  $H,r,\hat r_*$ (defined in
(\ref{standardcombinations})), and $\log\sbs[r]{\sT_{1,0}}$, $W_a$ are
computed from the Wronskian parameterization given in section
\ref{sec:wronskian}.  Secondly, $\hat h$ and
$\log\sbs[r]{Y_{1,1}Y_{2,2}}$ are found respectively from equations
(\ref{eq:hh}) and (\ref{dsYsa}). These auxiliary objects appear
in the source terms of (\ref{eq:39}), (\ref{eq:5}), (\ref{eq:7})\footnote{The perturbative expansion of
  (\ref{eq:7}), after subtracting the asymptotic solution is
  expressible through these auxiliary quantities,
  c.f. eight-loop approximations (\ref{eq:11}) and (\ref{appr1}).}, and
they turn out to be suppressed by one order of wrapping, hence their
contribution to the solution of the FiNLIE at given order is obtained
simply by knowing the solution of the FiNLIE at lower orders.

The position of the exact Bethe root constraints the solution through
the conditions $\sT_{1,0}^+(u_1)=0$ and $\sT_{1,1}(u_1)=0$. 
Let us stress that the ``asymptotic solutions''
(\ref{eq:as1}),(\ref{as:2}) are defined using the exact Bethe root
$u_1$, as opposed to the solution $\tilde u_1$ of the asymptotic
Beisert-Staudacher Bethe equation. This exact Bethe root is found from
equation (\ref{eq:BetheAn}). %
The correction $u_1-\tilde u_1$ to the position of the Bethe root is
suppressed by wrapping and, again, it can be computed iteratively. 

\subsection{\label{sec:analytic}Analytical structure of functions}
 It might  be not immediately clear that the equations presented above
 can be solved analytically at each order of the weak coupling
  expansion. We are going to show that this is indeed so
 and we will precise the class of functions which appear during this
 expansion. 

Let us start with the integration of densities $\rho$ and $\rho_2$, 
with either a Cauchy kernel $\CK$, like in \eqref{cT} and
\eqref{q2param}, or a Zhukovsky kernel like in \eqref{eq:6}. As we saw,
these densities are $\sqrt{1-z^2}$  times functions which are, at each
order in perturbation theory, polynomials in $z$. Actually, in the double
scaling regime (when $z=u/2g$ is fixed while $u\ll1$), which one has
to use for integrands if they are integrated on the finite support
$[-2g,2g]$,  any quantity 
 has an expansion which is either a
 polynomial or a polynomial times square root (or the sum of these two
 cases). Such expressions can be explicitly convoluted against both
 the Cauchy and the Zhukovsky kernel. The result of integration is
 either a polynomial in $z$, if the value of integral is computed in
 the double scaling regime, or a rational function in $u$, if it is to
 be computed in the ordinary regime, i.e with $g\to 0$ keeping $u$ fixed.

Let us now consider the weak coupling expansion of the quantities in
the ordinary regime.  It is easy to check that when expanding
asymptotic quantities in this regime, one always gets expressions
which can be represented as linear 
combinations of terms of the type $\bar\eta_{I}^{[-2]}\frac
1{(u-v)^a}\eta_{J}^{[2]}$. In the following such
linear combination will be called a standard-type expression. We will
now discuss two  non-trivial operations one encounters during solution
of the FiNLIE at weak coupling and show that both of them keep us in
the class of standard-type expressions\footnote{We do not discuss less
  complicated operations like algebraic manipulations, under which
  this class of functions is clearly stable.}. Hence, in this way we
will iteratively demonstrate that exact quantities are explicitly
computable and that they are always expressible in terms of
standard-type 
expressions.    

 The first operation is integration.  Except for the integrals over
 the finite support $[-2g,2g]$, which we discussed already, all
the integrals in the weak coupling version of the  FiNLIE have an
integration contour  from $-\infty$ to $+\infty$, parallel to the 
real axis. In section  \ref{sec:integration} and in  \appendixref \ref{sec:other-properties-eta} we showed how to integrate standard-type expression along such contour,
and the result is always a standard type expression. It is easy to
iteratively verify, starting from the asymptotic solution, that the
integrands are always of the standard type, and probably the only
troublesome place is the equation on $\hat h$, and more precisely the 
integration with $\coth(\pi(2gw-v))$. For this, we can note that 
\begin{equation}\label{cothexpr}
\coth(\pi(2gw-v)=-(\eta_1^{[2]}(v-2gw)+\bar\eta_1(u-2gw)).
\end{equation}
Hence, this $\coth$ is expressed as a linear combination of
$\eta$-functions at each order of expansion in $g^2$. Since 
$\eta$-functions obey the stuffle algebra and hence form a ring, we
see that at a given order in $g^2$, multiplying (\ref{cothexpr}) by a
standard-type expression always gives a standard-type expression.

The second operation is a semi-infinite summation.  One such summation is needed to
compute $U$: if $f=\log U/U^{[2]}$ is known from (\ref{eq:5})
then $\log U=\sum_{k=0}^\infty f^{[2k]}$. The other  semi-infinite sums
are used to compute $q_{13},q_{14},q_{24}$ according to
(\ref{eq:qij}). In both of these cases, the sums
are of the type $\sum_{k=0}^\infty f^{[2k]}$ for some function $f$. 
One can already note that,  in these sums, the corresponding functions
$f$ are analytic in the upper half-plane, hence the sums can involve
$\eta$-functions, but they should be free of any $\bar\eta$. Now, if
$f$-s contain only poles at positions $u=-\i\,k$, then the sum reduces
to a standard-type expression, e.g 
\begin{equation}
\sum_{k=0}^\infty \frac 1{(u+\i\, k)^a}\eta_{I}^{[2k+2]}=\eta_{a,I}.
\end{equation}
However, if there is a pole in another position, a more generic class
of functions (\ref{eq:9}) may  appear. In principle, one could have poles at Bethe
roots, but   we observe a remarkable cancellation of  these poles
for the physical solutions of the Y-system. 

To give an example of how the cancellation comes out, let us study the
computation of $q_{13}$ at the leading order, i.e. for
$U^2=-\frac{2g^4}{u^2}$ and $q_{12}=Q=u^2-u_1^2$, where
$u_1=-u_2=\frac 1 {2\sqrt 3}$. 
\begin{align}
q_{13}&=-2g^4Q\sum_{k=0}^\infty\left(\frac{1}{u^2Q^+Q^-}\right)^{[2k+1]}\nonumber\\
&=-18g^4Q\sum_{k=0}^\infty\left(\frac{1}{u^2}+\frac \i{2}\sum_{j=1}^2\left(\frac 1{u-\frac \i{2}-u_j}-\frac 1{u+\frac \i{2}+u_j}\right)\right)^{[2k+1]}
\nonumber\\&=-18g^4(\i\,u+Q\,\eta_2^+)\,.
\end{align}
We see that each pole at a position $u=u_j+\i\,k$ appears in two
successive terms of the sum which cancel each other. We checked this
cancellation mechanism at the first five orders of the perturbative expansion
(which was needed for computing the eight-loop anomalous
dimension). Hence we conclude that at least at these orders the
following equation, which is behind the cancellation mechanism, holds:    
\begin{align}\label{asBS2}
\left(\frac{\hat U^+}{\hat
    U^-}\right)^2=&-\frac{q_{12}^{[+2]}}{q_{12}^{[-2]}} 
\end{align}
at the zeroes of $q_{12}$ which are $u=u_{1}+\alpha$ and
$u=-u_1-\bar\alpha$. At the first four orders, we have $q_{12}=Q$ and
(\ref{asBS2}) is nothing but the asymptotic Bethe equation. At least
at the fifth order, (\ref{asBS2}) is still true, 
though it is no longer the asymptotic Bethe  equation. It is not the
exact Bethe equation either, although it is equivalent to it because
it does not follow from the constrains
$\sT_{1,0}^+(u_1)=\sT_{1,1}(u_1)=0$ and it is satisfied only if $u_1$
is the exact Bethe root. 

In order to have the same cancellation of poles in the sums expressing
$q_{14}$ and $q_{24}$, one should also require that 
\begin{align}\label{eq:q2}
\hat q_{2}^+=&\hat q_{2}^-&\textrm{at
  points }u\in&\{u_1+\alpha,-u_1-\bar\alpha\}\,.
\end{align}
In view of (\ref{eq:19b}), this requirement can be considered as a
regularity condition on $q_\emptyset$. At first four orders it is
equivalent to the equation $\sT_{1,1}(u_1)=0$, and we also verified
perturbatively that (\ref{eq:q2}) holds at the fifth order.

Under the
assumption that (\ref{asBS2}) and (\ref{eq:q2}) hold at any order of
the perturbation theory, all semi-infinite sums for $q_{ij}$ result in
standard-type expressions only.

By analyzing how $\hat H$ and $\hat r$ appears in the r.h.s. of
equation (\ref{eq:5}) for $U$, we see 
that the cancellation of poles in $q_{ij}$ propagates to the statement
that $\log U/U^{[2]}$ only has poles at position $u=-\i\,k$, hence $U$
is given by an expression of standard type. 

Let us stress that we verified (\ref{asBS2}) and (\ref{eq:q2}) at
the first five orders, and then predicted the general form of the
functions under the assumption that they hold at any order. These two
equations look very natural, since they ensure a more regular
structure of $q$-functions, by cancelling out some ladders of poles. If they
are not satisfied, then arbitrary generalized $\eta$-functions
 \eqref{eq:9} would appear. This class of
functions still forms a ring which insures that at most generalized
$\eta$-functions are present in the answer at any order; however the
answer would be considerably more complicated. 

This cancellation of poles at shifted Bethe roots does not mean that
$\eta$-functions {\it evaluated} at Bethe roots never appear. Such
numerical constants appear in various places, e.g. in the constraints
$\sT_{1,0}^+(u_1)=0$ and $\sT_{1,1}(u_1)=0$. For instance, they are
already present  in the leading displacement of the Bethe root. The
statement that we discussed above is merely
that there  are no $\eta$-functions that
depend {\it simultaneously} on the Bethe root and on the spectral
parameter. 

Since the computation of energy using  (\ref{wrapen}) reduces to the
computation of  integrals  of standard-type expressions, we 
conclude that if the 
constraints (\ref{asBS2}) and (\ref{eq:q2}) hold, then the perturbative
expansion of the energy is always given only by MZV-s and
by $\eta$-functions evaluated at Bethe roots. However,  $\eta$-functions
at Bethe roots are expected to cancel \cite{Bajnok:2008bm}. We observe
that they cancel indeed, 
at least up to  eight loops.

\section{\label{sec:energy}Anomalous dimension}
The anomalous dimension $\gamma_{\rm Konishi}=\Delta_{\rm Konishi}-4$ is
found from $\Delta_{\rm Konishi}=\Delta_{\rm as}(u_1)+\Delta_{\rm wrap}$, where
\begin{align}
\hspace{-2em}
\Delta_{\rm as}(u_1)=&4-8\,g\,{\rm Im}\left(\frac 1{\hat x_1^+}\right)\!,&
\label{wrapen}
        \Delta_{\rm wrap}=&\int_{\mathbb{R}-i0}\frac {-H(u)du}{\pi\sqrt{1-\frac{4g^2}{u^2}}}\,.
\end{align}
  As compared to the prediction from the asymptotic Bethe Ansatz,
  $\Delta_{\rm as}$ receives the following corrections through the
  displacement of the Bethe roots: 
\begin{align}
   \Delta_{\rm as}(u_1)= &\Delta_{\rm as}(\tilde u_{1})+\frac{4\i}{\sqrt{3}}g^{10}(1-5g^2+14g^4-461g^6)\sum_{k=1}^{4}m_kg^{2k-2}+\CO(g^{18}).
 \end{align}
The perturbative solution of the asymptotic Bethe equation gives
\begin{align}
\Delta_{\rm as}(\tilde u_{1})=&4+12\,g^{2}-48\,g^{4}+336\,g^{6}-12\,g^{8}(235+24\,\zeta_{3})
\no\\
&+12\,g^{10}(2209+360\,\zeta_{3}+240\,\zeta_{5})
\no\\
&-12\,g^{12}(22429+4608\,\zeta_{3}+3672\,\zeta_{5}+2520\,\zeta_{7})
\no\\
&+24\,g^{14}(119885+29064\,\zeta_{3}+144\,\zeta_{3}^{2}+24156\,\zeta_{5}
\no\\
&\hspace{6cm} +
19656\,\zeta_{7}+14112\,\zeta_{9})
\no\\
&-12\,g^{16}(2654761+742680\,\zeta_{3}+5760\,\zeta_{3}\,\zeta_{5}+6624\,\zeta_{3}^{2}+623904\,\zeta_{5}+
\no\\
&\qquad\qquad528552\,\zeta_{7}+447552\,\zeta_{9}+332640\,\zeta_{11})\,.
\end{align}
Separately $\Delta_{\rm as}(u_1)$ and $\Delta_{\rm wrap}$ depend on
$\eta$-functions evaluated at Bethe roots, however this dependence
cancels out   in their sum and one gets the result which only involves
Euler-Zagier sums: \PushButton[name=clickA2,
onclick={app.alert('4+12g^2-48g^4+336g^6+96g^8(-26+6
  zeta[3]-15zeta[5])
  -96g^10(-158-72zeta[3]+54zeta[3]^2+90zeta[5]-315zeta[7])
  -48g^12(160+5472zeta[3]-3240zeta[3]zeta[5]+432zeta[3]^2-2340zeta[5]-1575zeta[7]+10206zeta[9])
  +48g^14(-44480+108960zeta[3]+8568zeta[3]zeta[5]-40320zeta[3]zeta[7]-8784zeta[3]^2+2592zeta[3]^3
  -4776zeta[5]-20700zeta[5]^2-26145zeta[7]-17406zeta[9]+152460zeta[11])
  +48g^16(1133504+263736zeta[2]zeta[9]-1739520zeta[3]-90720zeta[3]zeta[5]-129780zeta[3]zeta[7]
  +78408zeta[3]zeta[8]+483840zeta[3]zeta[9]+165312zeta[3]^2-82080zeta[3]^2zeta[5]+41472zeta[3]^3
  +178200zeta[4]zeta[7]-409968zeta[5]+121176zeta[5]zeta[6]+463680zeta[5]zeta[7]+49680zeta[5]^2
  +455598zeta[7]+194328zeta[9]-555291zeta[11]-2208492zeta[13]-14256zeta[1,2,8])//Function[E,
  Symbol[FromCharacterCode[{916},"unicode"] ]==Series[E,{g,0,17}]]  
  ',3)}
]{\#}
\begin{flalign}
\label{konishianswer}
&\Delta_{\rm Konishi}=
4+12\,g^{2}-48\,g^{4}+336\,g^{6}+96\,g^{8}(-26+6\,\zeta_{3}-15\,\zeta_{5})
\nonumber\\
&\hspace{2em}-96\,g^{10}(-158-72\,\zeta_{3}+54\,\zeta_{3}^{2}+90\,\zeta_{5}-315\,\zeta_{7})
\nonumber\\
&\hspace{2em}-48\,g^{12}(160+5472\,\zeta_{3}-3240\,\zeta_{3}\,\zeta_{5}+432\,\zeta_{3}^{2}
\nonumber\\
&\hspace{7.5cm}
-2340\,\zeta_{5}-1575\,\zeta_{7}+10206\,\zeta_{9})
\nonumber\\
&\hspace{2em}+48\,g^{14}(-44480+108960\,\zeta_{3}+8568\,\zeta_{3}\,\zeta_{5}-40320\,\zeta_{3}\,\zeta_{7}-8784\,\zeta_{3}^{2}
\nonumber\\
&\hspace{0em}\hspace{3.7em}+2592\,\zeta_{3}^{3}-4776\,\zeta_{5}-20700\,\zeta_{5}^{2}-26145\,\zeta_{7}-17406\,\zeta_{9}+152460\,\zeta_{11})
\nonumber\\
&\hspace{2em}+48\,g^{16}(1133504+263736\,\zeta_{2}\,\zeta_{9}-1739520\,\zeta_{3}-90720\,\zeta_{3}\,\zeta_{5}
\nonumber\\
&\hspace{3.7em}-129780\,\zeta_{3}\,\zeta_{7}+78408\,\zeta_{3}\,\zeta_{8}+483840\,\zeta_{3}\,\zeta_{9}+165312\,\zeta_{3}^{2}
\nonumber\\
&\hspace{3.7em}-82080\,\zeta_{3}^{2}\,\zeta_{5}+41472\,\zeta_{3}^{3}+178200\,\zeta_{4}\,\zeta_{7}-409968\,\zeta_{5}+121176\,\zeta_{5}\,\zeta_{6}
\nonumber\\
&\hspace{3.7em}+463680\,\zeta_{5}\,\zeta_{7}+49680\,\zeta_{5}^{2}
+455598\,\zeta_{7}+194328\,\zeta_{9}-555291\,\zeta_{11}
\nonumber\\
&\hspace{7.5 cm}
-2208492\,\zeta_{13}-14256\,\zeta_{1,2,8})
\no\\
&\hspace{2em}+\CO(g^{18})\,.
\end{flalign}

We will now discuss some checks of the consistency of this result. In
the next paragraphs we will discuss the order of magnitude of the
answer, and then show to what extent this answer can be predicted from the
existing numeric data that gives the anomalous dimension for various values of the
coupling. Another check will be given in the next section, where
the terms having the highest transcendentality are derived.

The coefficients of the expansion \eqref{konishianswer} evaluate numerically to
\begin{multline}\label{eq:num}
\Delta_{\rm Konishi}\simeq
4+0.75000\,(4g)^{2}-0.18750\,(4g)^{4}+0.08203\,(4g)^{6}-0.05030\,(4g)^{8}\\
\hspace{2.3cm}
+0.03578\,(4g)^{10}-0.02728\,(4g)^{12}+0.02175(4g)^{14}-0.01791\,(4g)^{16}\,.\\[-.5cm]
\end{multline}
It is expected (see e.g. section 3 of \cite{Volin:2008kd}) that the
radius of convergency of  weak coupling expansions in the AdS/CFT
integrable system is at most $1/4$, because various Zhukovsky 
branch points in the $u$-plane collide when $g=\pm\i/4$
\footnote{There might be  other types of 
  singularities which depend on quantum numbers. For instance, the
  radius of convergency shrinks to zero when the spin quantum number
  approaches  the value $S=-1$.}. The coefficients in (\ref{eq:num})
are in agreement with this expectation.

One can further verify our results numerically using the  5-digit
precision data for the anomalous dimension \cite{Gromov:comm} which is given in figure \ref{fig:1}.
\begin{figure}
  \begin{center}
    \begin{tabular}{|c|c||c|c||c|c|}\hline
      g & $\Delta-4$ & g & $\Delta-4$ & g &$\Delta-4$ \\\hline
      0.23&0.53391(8)&0.45&1.4940(6)&0.71&2.5700(5)\\\hline 0.24&0.57395(9)&0.52&1.7987(8)&0.73&2.6464(1)\\\hline 0.25&0.61472(0)&0.53&1.8414(9)&0.75&2.7219(2)\\\hline 0.33&0.95885(9)&0.54&1.8839(8)&0.83&3.016(0)\\\hline 0.34&1.0032(8)&0.55&1.9262(1)&0.91&3.2988(2)\\\hline 0.35&1.0478(4)&0.61&2.174(7)&0.93&3.3679(9)\\\hline 0.43&1.4053(9)&0.63&2.2556(2)&0.96&3.4707(7)\\\hline 0.44&1.4497(9)&0.65&2.3356(2)&0.98&3.538(6)\\\hline
    \end{tabular}
  \end{center}
\caption{Numeric data provided by N.Gromov \cite{Gromov:comm} for the
  anomalous dimension of  the Konishi operator at various values of
  the coupling $g$.
\label{fig:1}
\PushButton[name=clickB, 
onclick={app.alert('{{23/100, 0.533918}, {6/25, 0.573959}, {1/4, 0.61472}, {33/100, 
  0.958859}, {17/50, 1.00328}, {7/20, 1.04784}, {43/100, 
  1.40539}, {11/25, 1.44979}, {9/20, 1.49406}, {13/25, 1.79878}, {53/
  100, 1.84149}, {27/50, 1.88398}, {11/20, 1.92621}, {61/100, 
  2.1747}, {63/100, 2.25562}, {13/20, 2.33562}, {71/100, 
  2.57005}, {73/100, 2.64641}, {3/4, 2.72192}, {83/100, 3.016}, {22/
  25, 3.1939}, {91/100, 3.29882}, {93/100, 3.36799}, {24/25, 
  3.47077}, {49/50, 3.5386}}
//Function[E,MatrixForm@N@Prepend[E,{"g",FromCharacterCode[{916, 45, 52},"unicode"]}]]
',3)}
]{\#}}
\end{figure}
 A Pad\'e approximant of a function usually has a larger radius of
 convergency than its Taylor series. And indeed, we empirically
 observe that the  following Pad\'e approximant 
\begin{equation}
        \Delta_{\rm Konishi}=4+\frac{12\,g^2+\sum\limits_{k=1}^{\Lambda} c_k\,g^{2k}}{1+\sum\limits_{k=1}^{\Lambda}d_k \,g^{2k}}
\end{equation}
converges, when increasing $\Lambda$, at
least for $g<0.6$. We took $\Lambda=10$,  expressed the coefficients
$c_1$, $c_2$, \ldots 
,$c_6$ through other coefficients so as to reproduce exactly the power
series expansion up to seven loops, and then we  fitted the the
remaining  $c_i$-s and $d_i$-s against the numerical data for $g<0.6$.
With this procedure, we got a prediction $0.01790\times 4^{16}$ for
the eight-loop coefficient of the series expansion, which is in perfect agreement with our
analytical result (\ref{konishianswer})\footnote{The validity of this approach can be tested
  by using it to compute 6- and 7-loop quantities
  first.}$^,$\footnote{In fact, one computes rather $\Delta_{\rm
    Konishi}-\Delta_{\rm as}(\tilde u_1)$ than $\Delta_{\rm Konishi}$. The
  magnitude of this difference is about 13\% of $\Delta_{\rm 
Konishi}$, hence the non-triviality of our test is in reproducing the first two digits of the
difference.}. 

\paragraph{Relation to knot numbers} There is a striking independent check of the result \eqref{konishianswer}, as we learned few months after the first preprint submission
of this article. As explained in \cite{Schnetz}, the large class Feynman graphs is evaluated in terms of a subclass of MZV-s: the so called single-valued MZV-s. Up to transcendentality 10 this subclass includes only single-indexed sums of odd argument, whereas at transcendentality 11 a new possible combination appears: $\zeta_{3,5,3}-\zeta_{3,5}\zeta_{3}$, which was observed in \cite{Broadhurst:1995km} at seven loops of $\phi^4$ theory.

Delightfully, our result is expressed in terms of single-valued MZV. \linebreak[4]Whereas the statement is obvious up to 7 loops, for the transcendentality $11$ piece of the $8$-loop term one  applies stuffle and shuffle relations to show that can be equivalently written as
\be
\ldots+g^{16}\,\frac{864}{5}\left(76307\,\zeta_{11}+792\,(\zeta_{3,5,3}-\zeta_{3,5}\,\zeta_3)-18840\,\zeta_3^2\,\zeta_5\right)+\ldots=\no\\
\ldots+g^{16}\,1728\,(132K_{3,5,3}-1752\,\zeta_3^2\,\zeta_5+7403\,\zeta_{11})+\ldots\,,
\ee
where $K_{3,5,3}$ is a knot number \cite{Broadhurst:1995km}.

\section{\label{sec:tran}Expansion in inverse transcendentality}
A good illustration of the approach explained in section \ref{sec:weak} is to perform an
expansion in inverse powers of transcendentality. Though this
expansion is technically more complicated, there are number of
benefits as well. In particular, the less  number of iterations is needed to capture
interesting higher-wrapping effects.  There is also no need to
distinguish the ordinary and double scaling regimes, and for the few leading orders
of expansion, only quite little should be known about the position of Bethe roots. 

The  strategy of this expansion is the following. We introduce a variable $\tau$, which
will be a bookkeeping variable 
for the inverse transcendentality order (the transcendentality
deficit). We assign this variable to the various quantities according
to the following rule\begin{equation}\label{taudistr} 
\begin{array}{r@{\;\;\leadsto\;\;}l}
g^a & \tau^{a} g^a\,,\\
\eta_{a_1,\ldots,a_k}^{[r]}& \tau^{-\sum_i a_i} \eta_{a_1,\ldots,a_k}^{[r]}\,,\\
\zeta_{a_1,\ldots,a_k}& \tau^{-\sum_i a_i} \zeta_{a_1,\ldots,a_k}\,,\\
\hat x, x,z & \phantom{\tau^0} \hat x, x,z\,,\\
(u+r)^{-a},\ a>0&\tau^{-a} (u+r)^{-a}\,.
\end{array}
\end{equation}
 After $\tau$ is assigned, we perform  an expansion in  powers of
 $\tau$. In most cases $\tau$ can be treated as an ordinary
 variable. However, there are exceptions: for instance, when one
 multiplies $(u+r_1)^{-a_1}$ and $(u+r_2)^{-a_2}$ with $r_1\neq r_2$,
 one should perform a partial fraction decomposition (i.e. rewrite the
 product as a sum of terms with a single pole), giving rise to a sum
 of terms with different orders in $\tau$, in which the leading term is of
 order $\tau^{-\max(a_1,a_2)}$. 
The transcendentality of $u$ in a numerator depends on whether it
cancels or not some  poles. For us, the latter issue is relevant only
for $q_2=-\i\, u+\ldots$, and in this case one can assign $\tau^{+1}$
for $u$. Another example of such exceptions is that the integration
parallel to the  real axis may increase the leading power of $\tau$ by
1 and to produce an expression with mixed powers of $\tau$.

To compute the leading transcendentality coefficients in the anomalous dimension to
all orders in $g$, let us assume that the leading transcendentality
$q$-functions can be derived solely from the large volume
solution. This assumption is justified by a careful analysis of
equations which can be found in \verb#Tanscendentality.nb#
\cite{Volin:link}.  

Under this assumption, we %
get
\begin{align}
       q_{12}=&Q+\CO(\tau^1),&q_2&=-\frac {\i}{3\,g\,x}\,\tau^{-1}+\CO(\tau^0),&U^2=&-\frac{2\,g^2}{ x^2}\tau^2+\CO(\tau^3)\,.
\end{align}
The leading transcendentality term for $q_{13}$ is then found from
\begin{align}
        q_{13}=&-2g^2\tau^2Q\,\sum_{n=0}^{\infty}\left(\frac 1{x^2Q^+Q^-}\right)^{[2n+1]}+\CO(\tau^3)\,.
\end{align}
In this sum, the highest transcendentality term is
determined as follows: at each order of $g$, we
write $\frac 1{x^2Q^+Q^-}$ as a sum of poles $\sum_k \frac
{\alpha_k} {\left(u-v_k\right)^{n_k}}$, so that the sum over $n$
simplifies to $\sum_k \eta_{n_k}^+(u-v_k)$, where the most
transcendental term comes from the pole with the largest
exponent. This is the pole coming from $\frac 1{x^2}$, and its
prefactor is the value of $\frac 1{Q^+Q^-}$ at $u=0$, namely
$9+\CO(g^2)$. Here it was enough to know the leading term in the
position of the Bethe root $u_1=\frac 1{\sqrt{12}}+\ldots$, because
the next terms are suppressed in $g$ and hence in $\tau$. We  see that
we  need to know quite little about the position of Bethe roots in
order to compute the leading transcendentality contributions.

Let us define, in analogy with (\ref{etafunctions})\footnote{The
  marginally divergent quantities are defined by
  $\hat\eta_1\equiv\sum(\frac 1{\hat x^{[2n]}}-\frac
  g{u+\i\,n})+\eta_1$ and then by stuffle relations.}, 
\begin{equation}\label{etahfunctions}
\hat\eta_{a_1,a_2,\ldots,a_k}\equiv \sum_{0\leq n_1<n_2<\ldots<n_k<\infty}\frac
1{(x^{[2n_1]})^{a_1}(x^{[2n_2]})^{a_2}\ldots (x^{[2n_k]})^{a_k}}\,.
\end{equation}
Note that $\hat\eta_I$ has the prefactor $\tau^0$.  For $q_{ij}$ one  gets then
\begin{align}
        q_{13}=&-18\,g^2\,\tau^2\,Q\,\hat\eta_2^+\,,&q_{14}=&6\, \i\, g\,\tau\,Q\,\hat\eta_2^+,&q_{24}=&2\,Q\,\hat\eta_4^+\,.
\end{align}
When we expand $H$ in orders of $\tau$, we encounter the  following
two combinations:
\begin{align}
        s_1=&\frac{\bar U^2}{Q^+}(q_{13}^+\,\bar q_2^2+2q_{14}^+\,\bar
        q_2+q_{24}^+),
& s_2=&\frac{q_{13}^+\, \bar q_{24}^-+2\,q_{14}^+\,\bar q_{14}^-+q_{24}^+\bar q_{13}^-}{Q^+Q^-}\,,
\end{align}
with $s_1\sim\tau^2$ and $s_2\sim \tau^2$. The expansion of $H$ in orders of $\tau$ gives
\begin{multline}\label{Hdw}
        H=\tau^2\frac{s_1}{Q^-Q^+}+\frac{\tau^4}{2}\left((9s_1)^2+18\,s_1s_2+\vphantom{\frac{(q_{1,4}^+)^2-q_{1,3}^+q_{1,4}^+}{(Q^+)^2}}\right.\\\left.18\,\bar U^2\frac{(q_{1,4}^+)^2-q_{1,3}^+q_{1,4}^+}{(Q^+)^2}\,\frac{\bar q_{13}^-\,\bar q_2^2-2\,\bar q_{14}^-\,\bar q_2+\bar q_{24}^-}{Q^-}\right)+\CO(\tau^5)\,,
      \end{multline}
     where the second term is a purely double wrapping effect. We see
     that at the leading order in $\tau$, the energy is given by the single wrapping term.

The above analysis allows computing the highest
transcendentality term to all orders in $g$. It is coming solely from $\Delta_{\rm wrap}$ in (\ref{wrapen}) and is given  by  the following explicit integral:
\begin{equation}
\Delta_{\rm lead\ tran}=2\i\times(-36\,g^2)\int\limits_{-\infty-\i\, 0}^{+\infty-\i\, 0}\frac{du}{-2\pi \i}\frac{1}{\sqrt{1-\frac{4g^2}{u^2}}}\left(\frac{\hat\eta_2^{[2]}}{\hat x^4}-\frac{2\,\hat\eta_3^{[2]}}{\hat x^3}+\frac{\hat\eta_4^{[2]}}{\hat x^2}\right)\,.
\end{equation}
To this end, we can use the Laplace representations
\begin{gather}
\frac{1}{\sqrt{1-\frac{4g^2}{u^2}}}\frac 1{\hat
  x^a}=\i^a\int_0^\infty dt\,e^{-\i\,u\,t}\,\partial_tJ_a(2\,g\,t),\\
\hat\eta_a^{[2]}=a\,(-\i)^a\int_0^\infty\frac{dt\,e^{\i\,u\,t}}{e^t-1}\frac{J_a(2\,g\,t)}{t}\,,
\end{gather}
which are simultaneously valid for   $0>\Im[u]>-1$. After some algebra, one gets
\begin{equation}\label{leadingtran}
\Delta_{\rm lead\ tran}=-432\,g^2\int_0^\infty \frac{d\,t}{e^t-1}\,\partial_t\left(\frac{J_3(2\,g\,t)^2}{t}\right)\,.
\end{equation}
In \verb#Tanscendentality.nb# \cite{Volin:link} we also performed a partial
analysis for subleading transcendentality terms, in particular for those
coming from the double wrapping term in (\ref{Hdw}), and
confirmed the coefficient of the $\zeta_{1,2,8}$ term in the anomalous
dimension at eight loops. 

\section{Conclusions and discussion}
In this paper we solved the AdS/CFT Y-system for the Konishi state
analytically at weak coupling up to the order where double wrapping
effects first appear. This allowed us  to compute the Konishi
anomalous dimension up to eight loops \eqref{konishianswer}. At this order we observed the
appearance of a non-reducible Euler-Zagier sum. In section
\ref{sec:analytic} we give a well justified prediction that the answer
will be a linear combination of Euler-Zagier sums at any order of
perturbation theory. In comparison, a superficial analysis of Feynman
graphs allows the appearance of non-reducible Euler-Zagier
sums \cite{Broadhurst:1996kc}, but it also allows
other types of numbers which may appear in the answer starting from
nine loops\footnote{We thank Matthias Staudacher for clarifications
  about this point.} \cite{Brown:2010bw}.

Our computation is based on the FiNLIE that was proposed in
\cite{Gromov:2011cx} and was first adjusted for the weak coupling expansion in
\cite{PhysRevLett.109.241601}.  We presented  all the integral
equations which allow one to straightforwardly perform the weak coupling
expansion to any desired order. By contrast, the approach based on
Luscher formulae is inapplicable beyond single-wrapping
orders\footnote{For vacuum state the double-wrapping Luscher formulae
  are known \cite{Ahn:2011xq}. It is not clear, however, how to
  generalize these formulae  to excited states.}.  

It appears that at any order
of the perturbative expansion,  the FiNLIE quantities are expressed 
as linear combinations of products of a
multiple Hurwitz zeta function times the complex conjugate of a
multiple Hurwitz zeta function, where the coefficients of the linear
combinations are rational functions. We showed that all the integrals
can be evaluated analytically in terms of this basis. We developed
{\it Mathematica} packages that handle the technical details  of the
computation and which were used to compute the leading double wrapping
order. It would be  possible  to apply the FiNLIE approach  for other
weak coupling computations, e.g for the angle-dependent cusp anomalous
dimension (c.f. \cite{Gromov:2012eu}) or the BMN vacuum of the
$\gamma$-deformed theory. We also believe that the {\it Mathematica}
tools which we provide here will be useful in a broader context.  

The main  obstacle for even higher-loop computations is combinatorial
growth of the basis of multiple Hurwitz zeta functions which leads to
an exponentially increasing demand of computing resources. We
estimate, however, that it is technically feasible to reach
triple-wrapping orders if a motivation arises for this feat. 

\ \\
An interesting observation is that the algebra of multiple Hurwitz
zeta functions appearing in the AdS/CFT integrability respects the
stuffle relations, while the algebra of polylogarithms, more typical
for perturbative quantum field theory, respects the shuffle
relations. Hence, in a sense the two algebras are complementary which
might be a hint  for a richer Hopf algebra structure of the planar
{\sym}. 

\ \\
Instead of performing an expansion in the coupling constant, it is
also possible to perform an expansion in inverse orders of
transcendentality. This approach captures some effects from higher
loops, and in particular we computed the most
transcendental terms to all loop orders (\ref{leadingtran}) in section
\ref{sec:tran}. Probably the higher transcendentality terms may also
be captured by perturbative quantum field theory methods, which would provide an
interesting venue for comparison. 

\ \\
There is another interesting physical phenomenon discussed  in section
\ref{sec:analytic}: The exact Bethe equations can be interpreted as a
regularity condition on the Y-system. This feature was already
questioned in \cite{Gromov:2011cx}, and its first confirmation was
obtained in \cite{PhysRevLett.109.241601}, but only for asymptotic
quantities. Here we demonstrated that the cancellation of poles in the
$q$-functions is ensured even if the leading wrapping correction is
taken into account. Quite interestingly, we observed a cancellation of
poles at the shifted  zeroes of $q_{12}$ and not at the shifted
positions of the Bethe roots $\pm u_1$.
 $q_{12}$ is equal to the Baxter polynomial $Q$ only asymptotically,
 but wrapping effects make it deviate from $Q$, and then its zeroes
 acquire a positive imaginary part. On the one hand, $q_{12}\neq Q$
 could be an artefact of the computation scheme.  On the other hand,
 Bethe roots are known to become complex in e.g. Lee-Yang model at
 finite volume \cite{Dorey:1996re}, hence one can speculate on another interesting
 possibility: the zeroes $q_{12}$ are an alternative way to
 parameterize the physical state. This is especially
 appealing because if we assume that these zeroes are determined by
 the regularity conditions, then  equation (\ref{asBS2}) should be exactly satisfied. On the other hand, this equation can be equivalently rewritten as
 an equation
\begin{equation}
        \frac{\hat q_{123}^{[+1]}\hat q_1^{[+1]}}{\hat q_{123}^{[-1]}\hat q_1^{[-1]}}=-\frac{\hat q_{12}^{[+2]}}{\hat q_{12}^{[-2]}}\,\
\ {\rm for\ \ } u - {\rm zero\ of\ } {\hat q}_{12}\,,
\end{equation}
which takes precisely the same form as the Bethe  equations appearing
in the algebraic Bethe Ansatz solution of spin chains. Hence the
AdS/CFT integrable system has one more  common feature with
integrable systems solvable by an algebraic Bethe Ansatz, in addition
to the group-theoretical interpretation of  
T-functions proposed in \cite{Gromov:2011cx}.

\ \\
{\bf Acknowledgments.} We thank Zoltan Bajnok, Vladimir Kazakov, Gregory Korchemsky, Oliver Schnetz, and
Matthias Staudacher for interesting discussions and Nikolay Gromov
for providing us numerical data. The work of S.L. is supported by
the ERC Advanced grant No.290456. S.L. thanks Nordita for hospitality
where a part of this work was done. 

\appendix
%

\section{Further details about \texorpdfstring{$\eta$}{eta}-functions}
\label{sec:other-properties-eta}

\subsection{Marginally divergent \texorpdfstring{$\eta$}{eta}-functions}
\label{sec:marg-diverg-eta}

Like in the case of MZVs, the sum \eqref{etafunctions} is convergent
under the conditions $a_k>1$, $a_k+a_{k-1}>2$, \ldots,
$\sum\limits_{i=1}^k a_i>k$. By contrast, in the case of 
$\eta_1$, the sum is logarithmically divergent, and we {\it define}
the regularized value of $\eta_1$ by (\ref{etapolygamma}).
We then {\it define} the regularized value of $\zeta_1$ by
(\ref{zetaeta}), and this definition gives $\zeta_1\equiv\gamma_{\rm
  Euler-Mascheroni}$. %

All  the marginally divergent $\eta$-functions have the form
$\eta_{a_1,\ldots,a_k,1,1,\ldots,1}$, with $a_k>1$ or $k=0$, and we
{\it define} them by requiring that stuffle algebra relations are
satisfied.  One can check that this definition is self-consistent.

 Explicitly, we have
 \begin{equation}
   \begin{aligned}
   \eta_{a_1,\ldots,a_k,1}\equiv&
   \eta_{1}\eta_{a_1,\ldots,a_k}-\sum_{i=1}^k\eta_{a_1,\ldots,a_{i-1},1,a_i,\ldots,a_k}-\sum_{i=1}^k\eta_{a_1,\ldots,a_{i-1},a_i+1,a_{i+1},\ldots,a_k}\,,\\
   \eta_{a_1,\ldots,a_k,1,1}\equiv& \frac 12\left(\eta_1\eta_{a_1,\ldots,a_k,1}-\sum_{i=1}^k\eta_{a_1,\ldots,a_{i-1},1,a_i,\ldots,a_k,1}
  \right.\\& \hspace{5em}\left.-\sum_{i=1}^k\eta_{a_1,\ldots,a_{i-1},a_i+1,a_{i+1},\ldots,a_k,1}-\eta_{a_1,\ldots,a_k,2}\,\right),\\ \cdots
   \end{aligned}
 \end{equation}
In this way, we recursively show that all
marginally divergent functions are  expressed as polynomials in
$\eta_1$ with coefficients which are linear in convergent
$\eta$-functions. In particular, there is an explicit formula 
\begin{equation}
        \eta_{\scriptsize\underbrace{1,1,\ldots,1}_{(\rm k\ times)}}=\frac 1{k!}(\eta_{1}+\partial_u)^k\eta_1\,.\label{eq:27}
\end{equation}
 Using these relations and (\ref{zetaeta}), one defines marginally divergent MZVs.

\subsection{Factorization property}
\label{sec:expr-some-gener}
As discussed in section \ref{sec:integration}, there is a simple
recursive way to express integrals of the form \eqref{eq:17}. After a
few iterations, any such integral is expressed in terms of standard
$\eta$-functions evaluated either at point $u=\i$ (where they are
equal to MZVs (\ref{zetaeta})) or at point $u=v$, and of generalized
$\eta$-functions of the form $\eta_K^{\{\ldots,0,0,v,0,0,\ldots\}}$,  
evaluated at point $u=\i$.
For instance, for the integral \eqref{eq:30}, this procedures gives
\begin{equation}
  \label{eq:38}
  \int_{-\infty}^{+\infty}\frac{du}{-2\pi
    \i}\,\bar\eta_{3}^{[-2]}\frac 1{(u-v)}\eta_{2}^{[2]}=\left.-3
    \eta_{1,4}^{\{v,0\}}-2\eta_{2,3}^{\{v,0\}}-\eta_{3,2}^{\{v,0\}}\right|_{u=\i}+
\left.\bar\eta_{3}^{[-2]}\eta_{2}^{[2]}\right|_{u=v}\,.
\end{equation}

 In the expression obtained by this method, it is quite
important that the generalized $\eta$-functions that appear have
only one non-trivial shift $v$. This actually allows us to express them in terms of
standard $\eta$-functions and $\bar\eta$-functions. For example, one
obtains \PushButton[name=click567,
onclick={app.alert('The result is evaluated from the following input: 
(GeneralizedEta[2, 2][{-u, 0}][I] /. EndSimplify //. shiftmeta[2] // FuntoMHZbasis)/.suboppositesign',3)}
]{\#}
\begin{align}
\left.\eta_{2,2}^{\{v,0\}}\right|_{u=\i}=&
\left. \frac {\zeta_2}{u^2}+2\,\i\, \zeta_1 \, \bar\eta_3 - \bar\eta_{2,2} -2 \,\bar\eta_{3,1}
\right|_{u=v}\,.
\end{align}
The proof of this  factorization property is done by induction over the
depth of the $\eta$-functions. First, it is clear that
$\left.\eta_{a}^{\{v\}}\right|_{u=\i}=\eta_a(\i-v)=(-1)^{a}\left.\bar\eta^{[-2]}_a\right|_{u=v}$. Let
us now  assume that, when 
the depth  is smaller than $n$, we know 
how to express $\left.\eta_K^{\{\ldots,0,0,v,0,0,\ldots\}}\right|_{u=i}$ in terms of
standard $\eta$-functions and $\bar\eta$-functions. Then, for an arbitrary multi-index
$I=a,\check I$ with
$n$ elements, we have two different ways to compute the integral
\begin{align}
  \label{eq:20}
  \int_{-\infty}^{+\infty}\frac{du}{-2\pi
    \i}\, \frac {\bar \eta _{\check I,1}^{[-2]}}{\left(u-v\right)^a}
\end{align}
for  $0<{\rm Im}(v)<1$: either we close the integration contour
downwards, and we immediately 
obtain that this integral is zero. Or we close it upwards,
using the iterative procedure from above. Except for the very last
residue, at each step  we will obtain expressions involving
$\eta$-functions with generalized shifts of the form 
$\eta_K^{\{\ldots,0,0,v,0,0,\ldots\}}$, where $K$ has less than $n$
elements. On the other hand, if we denote  $\tilde
I=a_n,a_{n-1},\ldots,a_1$ (where $I=a_1,a_2,\ldots,a_n$), then the
residue at the very last step is
$\res_{u=0}\frac{\left(\bar \eta_{\tilde I}^{\{
\ldots 0,0,v
\}}\right)^{[+2]}}{u}=\left.\bar \eta_{\tilde I}^{\{
\ldots 0,0,v
\}}\right|_{u=i}$.
 Hence the vanishing of the integral \eqref{eq:20}
allows us to express $\left.\bar \eta_{\tilde I}^{\{
\ldots 0,0,v
\}}\right|_{u=i}$ in terms of usual $\eta$-functions.
For instance, this argument allows to derive the relation
\begin{equation}
  \label{eq:31}
  \left.\bar\eta_{2,2}^{\{0,v\}}\right|_{u=\i}=
\left.  -\zeta_2\eta_2^{[2]}
  +2\,\i\,\zeta_1\eta_3^{[2]}+\eta_{2,2}^{[2]}+2 \,\eta_{3,1}^{[2]}
\right|_{u=v}
\,.
\end{equation}
 Finally, one notes that the
functions  $\eta_K^{\{\ldots,0,0,v,0,0,\ldots\}}$ where the  shift $v$
is not in the last position are
expressed through the functions $\eta_{\tilde K}^{\{\ldots,0,0,v\}}$
using the stuffle algebra. For instance, we have 
\begin{align}
  \label{eq:21}
  \eta_{a,b,c}^{\{0,v,0\}} =&  \eta_{a,b}^{\{0,v\}}   \eta_{c}%
-  \eta_{a,c,b}^{\{0,0,v\}} -\eta_{c,a,b}^{\{0,0,v\}} + \cdots\,,
\end{align}
 where the period ($\cdots$) denotes  $\eta$-functions with less indices. 

Hence, we showed how to express any basic integral (\ref{eq:17}) in
terms of $\eta$- and $\bar\eta$-functions, MZVs, and rational
functions of $u$. 

It is clear that if $\Im(v)>0,$ the answer for the integral  (\ref{eq:17}) should
be analytic in the upper half-plane. We will now describe relations
that allow canceling all the $\bar\eta$-functions using the
periodicity property, and these relations can be used to make the
analyticity of integrals of the form (\ref{eq:17}) manifest.

\subsection{Periodicity property}
\label{sec:periodicity}
There is
a certain class of 
relations between $\bar\eta$- and $\eta$-functions. One example of
such relation is: 
\begin{align}
  \label{eq:8}
\left(\eta _2^{[+2]}+\bar{\eta }_2^{}\right)^2=%
-4 \zeta _2 \left(\eta _2^{[+2]}+\bar{\eta }_2^{}\right)+\left(\eta _4^{[+2]}+\bar{\eta }_4^{}\right)\,.
\end{align}

To prove the relation \eqref{eq:8}, we  notice that
the function $%
  (\eta _2^{[+2]}+\bar{\eta }_2^{}%
)^2
=%
  (\sum_{k\in \mathbb Z}\frac 1 {\left(u+i k %
  \right)^2 } %
)^2
$ is a periodic function with period $i$, which has the Laurent series
expansion $%
  (\eta _2^{[+2]}+\bar{\eta }_2^{}%
)^2
=\frac{1}{u^4}-\frac{4 \zeta _2}{u^2}+\frac {44} 5 \zeta _2^2+\mathcal{O}(u^2)$
in the vicinity of zero, which is its only singularity lying in the strip $\{u:
|\mathrm{Im}(u)|\leq 1/2\}$. In the r.h.s.
of \eqref{eq:8}, the coefficient %
$-4 \zeta _2$ is chosen in such a way
that the 
r.h.s., which
is also periodic with poles on $i \mathbb Z$, has the same Laurent series
in the vicinity of zero. Hence, since also both the l.h.s. and the
r.h.s.  tend to zero when $u\to \pm \infty$, they should be equal.

Using the same idea, we will actually show that
an arbitrary function $\bar \eta_I$ can be expressed in
terms of the functions $\eta_{J}$ and of the function $\bar
\eta_1$.
 To this end, let us study the periodic function %
 \begin{align}
   \mathcal{P}_{a,I}=&\sum_{k\in\mathbb Z} \frac 1 {\left(u+i
       k\right)^a} \eta_{I}\left(u+i (k+1)\right)\,.\label{eq:13}
 \end{align}
If $I$
is the empty multi-index, we use the convention
$\eta_{\emptyset}=1$, hence we have
\begin{align}
  \mathcal{P}_{a}=&\sum_{k\in\mathbb Z} \frac 1 {\left(u+i
      k\right)^a} = \eta_a^{[+2]} +  \bar \eta _a\,.
\label{eq:14}
\end{align}
Let us denote by $i_2,i_3,\ldots,i_d$ the elements of $I$, and define
$i_1=a$. Then one can generalize the relation \eqref{eq:14} as follows:
\begin{align}
  \label{eq:33}
  \mathcal{P}_{a,I}=& \sum_{n_1<n_2<\ldots<n_d} \left(\frac 1
    {u^{i_1}}\right)^{[2n_1]}
\left(\frac 1 {u^{i_2}}\right)^{[2n_2]} \cdots \left(\frac 1 {u^{i_d}}\right)^{[2n_d]}
 \\
=& \sum_{k=0}^d\quad \sum_{n_1<n_2<\ldots<n_k\leq 0<n_{k+1}<\ldots
  <n_d}\quad \prod_{p=1}^d
 \left(\frac 1
    {u^{i_p}}\right)^{[2n_p]}\\
=&\sum_{k=0}^d \, \eta_{i_{k+1},i_{k+2},\ldots,i_d}^{[2]} \,\,\bar \eta _{i_k,i_{k-1},\ldots,i_1}\,,\label{eq:35}
\end{align}
where we remind the convention $\eta_\emptyset=\bar\eta_\emptyset=1$. For
instance, if $I$ has two elements $b$ and $c$, then we have
\begin{align}
  \label{eq:34}
  \mathcal{P}_{a,b,c}=&\eta _{a,b,c}^{[2]}+\eta _{b,c}^{[2]} \bar{\eta }_a+\eta _c^{[2]} \bar{\eta }_{b,a}+\bar{\eta }_{c,b,a}\,.
\end{align}

Using the periodicity of $\mathcal{P}_I$, we can %
then
find a set of constants $\gamma_k$ such that $\mathcal{P}_I=
\sum_{k}\gamma_k\left( \eta_1^{[+2]}+\bar \eta_1\right)^k$: these
constants are found by requiring that the singular and constant parts
of the Laurent 
series of the r.h.s. and the l.h.s. do match. Using the expression
\eqref{eq:35} of
$\mathcal{P}_I$, this allows  writing 
$\bar \eta_{I} = 
\sum_{k}\gamma_k\left( \eta_1^{[+2]}+\bar \eta_1\right)^k -
\sum_{k=0}^{d-1} \, \eta_{i_{k+1},i_{k+2},\ldots,i_d}^{[2]} \,\,\bar \eta
_{i_k,i_{k-2},\ldots,i_1}$. %
We can therefore iteratively express an arbitrary function $\bar
\eta_{I}$ in terms of  functions $\eta_{J}$ and of the function $\bar
\eta_1$.


\bibliographystyle{elsarticle-num}
\bibliography{bibliography}

\begin{thebibliography}{10}
\expandafter\ifx\csname url\endcsname\relax
  \def\url#1{\texttt{#1}}\fi
\expandafter\ifx\csname urlprefix\endcsname\relax\def\urlprefix{URL }\fi
\expandafter\ifx\csname href\endcsname\relax
  \def\href#1#2{#2} \def\path#1{#1}\fi

\bibitem{Volin:link}
S.~Leurent, D.~Volin,
  \href{{http://people.kth.se/~dmytrov/konishi8.zip}}{\hspace{-9em}\raisebox{-%
.6em}{\parbox[c]{\textwidth}{\hspace{9.3em}{\it {\MakeUppercase{m}}athematica}
  packages for working with zeta\\ functions and
  {\MakeUppercase{f}}i{\MakeUppercase{nlie}}-based weak coupling expansion.}}}
\newline\urlprefix\url{{http://people.kth.se/~dmytrov/konishi8.zip}}

\bibitem{Beisert:2010jr}
N.~Beisert, C.~Ahn, L.~F. Alday, Z.~Bajnok, J.~M. Drummond, et~al., {Review of
  AdS/CFT Integrability: An Overview}, Lett.Math.Phys. 99 (2012) 3--32.
\newblock \href {http://arxiv.org/abs/1012.3982} {\path{arXiv:1012.3982}},
  \href {http://dx.doi.org/10.1007/s11005-011-0529-2}
  {\path{doi:10.1007/s11005-011-0529-2}}.

\bibitem{Ambjorn:2005wa}
J.~Ambjorn, R.~A. Janik, C.~Kristjansen, {Wrapping interactions and a new
  source of corrections to the spin-chain / string duality}, Nucl. Phys. B736
  (2006) 288--301.
\newblock \href {http://arxiv.org/abs/hep-th/0510171}
  {\path{arXiv:hep-th/0510171}}, \href
  {http://dx.doi.org/10.1016/j.nuclphysb.2005.12.007}
  {\path{doi:10.1016/j.nuclphysb.2005.12.007}}.

\bibitem{Beisert:2005fw}
N.~Beisert, M.~Staudacher, {Long-range $PSU(2,2|4)$ Bethe Ansaetze for gauge
  theory and strings}, Nucl. Phys. B727 (2005) 1--62.
\newblock \href {http://arxiv.org/abs/hep-th/0504190}
  {\path{arXiv:hep-th/0504190}}, \href
  {http://dx.doi.org/10.1016/j.nuclphysb.2005.06.038}
  {\path{doi:10.1016/j.nuclphysb.2005.06.038}}.

\bibitem{Beisert:2006ez}
N.~Beisert, B.~Eden, M.~Staudacher, {Transcendentality and Crossing},
  J.Stat.Mech. 0701 (2007) P01021.
\newblock \href {http://arxiv.org/abs/hep-th/0610251}
  {\path{arXiv:hep-th/0610251}}, \href
  {http://dx.doi.org/10.1088/1742-5468/2007/01/P01021}
  {\path{doi:10.1088/1742-5468/2007/01/P01021}}.

\bibitem{Gromov:2009tv}
N.~Gromov, V.~Kazakov, P.~Vieira, {Exact Spectrum of Anomalous Dimensions of
  Planar N=4 Supersymmetric Yang-Mills Theory}, Phys. Rev. Lett. 103 (2009)
  131601.
\newblock \href {http://arxiv.org/abs/0901.3753} {\path{arXiv:0901.3753}},
  \href {http://dx.doi.org/10.1103/PhysRevLett.103.131601}
  {\path{doi:10.1103/PhysRevLett.103.131601}}.

\bibitem{Gromov:2009zb}
N.~Gromov, V.~Kazakov, P.~Vieira, {Exact Spectrum of Planar ${\cal N}=4$
  Supersymmetric Yang- Mills Theory: Konishi Dimension at Any Coupling}, Phys.
  Rev. Lett. 104 (2010) 211601.
\newblock \href {http://arxiv.org/abs/0906.4240} {\path{arXiv:0906.4240}},
  \href {http://dx.doi.org/10.1103/PhysRevLett.104.211601}
  {\path{doi:10.1103/PhysRevLett.104.211601}}.

\bibitem{Frolov:2010wt}
S.~Frolov, {Konishi operator at intermediate coupling}, J. Phys. A44 (2011)
  065401.
\newblock \href {http://arxiv.org/abs/1006.5032} {\path{arXiv:1006.5032}},
  \href {http://dx.doi.org/10.1088/1751-8113/44/6/065401}
  {\path{doi:10.1088/1751-8113/44/6/065401}}.

\bibitem{Frolov:2012zv}
S.~Frolov, {Scaling dimensions from the mirror TBA}, J.Phys. A45 (2012) 305402.
\newblock \href {http://arxiv.org/abs/1201.2317} {\path{arXiv:1201.2317}},
  \href {http://dx.doi.org/10.1088/1751-8113/45/30/305402}
  {\path{doi:10.1088/1751-8113/45/30/305402}}.

\bibitem{Gromov:2011de}
N.~Gromov, D.~Serban, I.~Shenderovich, D.~Volin, {Quantum folded string and
  integrability: From finite size effects to Konishi dimension}, JHEP 1108
  (2011) 046.
\newblock \href {http://arxiv.org/abs/1102.1040} {\path{arXiv:1102.1040}},
  \href {http://dx.doi.org/10.1007/JHEP08(2011)046}
  {\path{doi:10.1007/JHEP08(2011)046}}.

\bibitem{Gromov:2011bz}
N.~Gromov, S.~Valatka, {Deeper Look into Short Strings}, JHEP 1203 (2012) 058.
\newblock \href {http://arxiv.org/abs/1109.6305} {\path{arXiv:1109.6305}},
  \href {http://dx.doi.org/10.1007/JHEP03(2012)058}
  {\path{doi:10.1007/JHEP03(2012)058}}.

\bibitem{Roiban:2009aa}
R.~Roiban, A.~A. Tseytlin, {Quantum strings in AdS$_5$ x S$^5$: strong-coupling
  corrections to dimension of Konishi operator}, JHEP 11 (2009) 013.
\newblock \href {http://arxiv.org/abs/0906.4294} {\path{arXiv:0906.4294}},
  \href {http://dx.doi.org/10.1088/1126-6708/2009/11/013}
  {\path{doi:10.1088/1126-6708/2009/11/013}}.

\bibitem{Roiban:2011fe}
R.~Roiban, A.~Tseytlin, {Semiclassical string computation of strong-coupling
  corrections to dimensions of operators in Konishi multiplet}, Nucl.Phys. B848
  (2011) 251--267.
\newblock \href {http://arxiv.org/abs/1102.1209} {\path{arXiv:1102.1209}},
  \href {http://dx.doi.org/10.1016/j.nuclphysb.2011.02.016}
  {\path{doi:10.1016/j.nuclphysb.2011.02.016}}.

\bibitem{Vallilo2011}
B.~C. Vallilo, L.~Mazzucato, {The Konishi multiplet at strong coupling},
  Journal of High Energy Physics 2011 (2011) 1--9.
\newblock \href {http://arxiv.org/abs/1102.1219} {\path{arXiv:1102.1219}},
  \href {http://dx.doi.org/10.1007/JHEP12(2011)029}
  {\path{doi:10.1007/JHEP12(2011)029}}.

\bibitem{Bajnok:2008bm}
Z.~Bajnok, R.~A. Janik, {Four-loop perturbative Konishi from strings and finite
  size effects for multiparticle states}, Nucl. Phys. B807 (2009) 625--650.
\newblock \href {http://arxiv.org/abs/0807.0399} {\path{arXiv:0807.0399}},
  \href {http://dx.doi.org/10.1016/j.nuclphysb.2008.08.020}
  {\path{doi:10.1016/j.nuclphysb.2008.08.020}}.

\bibitem{Bajnok:2009vm}
Z.~Bajnok, A.~Hegedus, R.~A. Janik, T.~Lukowski, {Five loop Konishi from
  AdS/CFT}, Nucl.Phys. B827 (2010) 426--456.
\newblock \href {http://arxiv.org/abs/0906.4062} {\path{arXiv:0906.4062}},
  \href {http://dx.doi.org/10.1016/j.nuclphysb.2009.10.015}
  {\path{doi:10.1016/j.nuclphysb.2009.10.015}}.

\bibitem{Arutyunov:2010gb}
G.~Arutyunov, S.~Frolov, R.~Suzuki, {Five-loop Konishi from the Mirror TBA},
  JHEP 04 (2010) 069.
\newblock \href {http://arxiv.org/abs/1002.1711} {\path{arXiv:1002.1711}},
  \href {http://dx.doi.org/10.1007/JHEP04(2010)069}
  {\path{doi:10.1007/JHEP04(2010)069}}.

\bibitem{Balog:2010xa}
J.~Balog, A.~Hegedus, {5-loop Konishi from linearized TBA and the XXX magnet},
  JHEP 06 (2010) 080.
\newblock \href {http://arxiv.org/abs/1002.4142} {\path{arXiv:1002.4142}},
  \href {http://dx.doi.org/10.1007/JHEP06(2010)080}
  {\path{doi:10.1007/JHEP06(2010)080}}.

\bibitem{Fiamberti:2007rj}
F.~Fiamberti, A.~Santambrogio, C.~Sieg, D.~Zanon, {Wrapping at four loops in
  N=4 SYM}, Phys.Lett. B666 (2008) 100--105.
\newblock \href {http://arxiv.org/abs/0712.3522} {\path{arXiv:0712.3522}},
  \href {http://dx.doi.org/10.1016/j.physletb.2008.06.061}
  {\path{doi:10.1016/j.physletb.2008.06.061}}.

\bibitem{Fiamberti:2008sh}
F.~Fiamberti, A.~Santambrogio, C.~Sieg, D.~Zanon, {Anomalous dimension with
  wrapping at four loops in N=4 SYM}, Nucl.Phys. B805 (2008) 231--266.
\newblock \href {http://arxiv.org/abs/0806.2095} {\path{arXiv:0806.2095}},
  \href {http://dx.doi.org/10.1016/j.nuclphysb.2008.07.014}
  {\path{doi:10.1016/j.nuclphysb.2008.07.014}}.

\bibitem{Velizhanin:2008jd}
V.~Velizhanin, {The four-loop anomalous dimension of the Konishi operator in
  N=4 supersymmetric Yang-Mills theory}, JETP Lett. 89 (2009) 6--9.
\newblock \href {http://arxiv.org/abs/0808.3832} {\path{arXiv:0808.3832}},
  \href {http://dx.doi.org/10.1134/S0021364009010020}
  {\path{doi:10.1134/S0021364009010020}}.

\bibitem{Eden:2012fe}
B.~Eden, P.~Heslop, G.~P. Korchemsky, V.~A. Smirnov, E.~Sokatchev, {Five-loop
  Konishi in N=4 SYM}, Nucl.Phys. B862 (2012) 123--166.
\newblock \href {http://arxiv.org/abs/1202.5733} {\path{arXiv:1202.5733}},
  \href {http://dx.doi.org/10.1016/j.nuclphysb.2012.04.015}
  {\path{doi:10.1016/j.nuclphysb.2012.04.015}}.

\bibitem{Bajnok:2008qj}
Z.~Bajnok, R.~A. Janik, T.~Lukowski, {Four loop twist two, BFKL, wrapping and
  strings}, Nucl. Phys. B816 (2009) 376--398.
\newblock \href {http://arxiv.org/abs/0811.4448} {\path{arXiv:0811.4448}},
  \href {http://dx.doi.org/10.1016/j.nuclphysb.2009.02.005}
  {\path{doi:10.1016/j.nuclphysb.2009.02.005}}.

\bibitem{Lukowski:2009ce}
T.~Lukowski, A.~Rej, V.~N. Velizhanin, {Five-Loop Anomalous Dimension of
  Twist-Two Operators}, Nucl. Phys. B831 (2010) 105--132.
\newblock \href {http://arxiv.org/abs/0912.1624} {\path{arXiv:0912.1624}},
  \href {http://dx.doi.org/10.1016/j.nuclphysb.2010.01.008}
  {\path{doi:10.1016/j.nuclphysb.2010.01.008}}.

\bibitem{Balog:2010vf}
J.~Balog, A.~Hegedus, {The Bajnok-Janik formula and wrapping corrections}, JHEP
  09 (2010) 107.
\newblock \href {http://arxiv.org/abs/1003.4303} {\path{arXiv:1003.4303}},
  \href {http://dx.doi.org/10.1007/JHEP09(2010)107}
  {\path{doi:10.1007/JHEP09(2010)107}}.

\bibitem{Kotikov:2007cy}
A.~V. Kotikov, L.~N. Lipatov, A.~Rej, M.~Staudacher, V.~N. Velizhanin,
  {Dressing and Wrapping}, J. Stat. Mech. 0710 (2007) P10003.
\newblock \href {http://arxiv.org/abs/0704.3586} {\path{arXiv:0704.3586}},
  \href {http://dx.doi.org/10.1088/1742-5468/2007/10/P10003}
  {\path{doi:10.1088/1742-5468/2007/10/P10003}}.

\bibitem{PhysRevLett.109.241601}
S.~Leurent, D.~Serban, D.~Volin, {Six-Loop Konishi Anomalous Dimension from the
  $Y$ System}, Phys. Rev. Lett. 109 (2012) 241601.
\newblock \href {http://arxiv.org/abs/1209.0749} {\path{arXiv:1209.0749}},
  \href {http://dx.doi.org/10.1103/PhysRevLett.109.241601}
  {\path{doi:10.1103/PhysRevLett.109.241601}}.

\bibitem{Bajnok:2012bz}
Z.~Bajnok, R.~A. Janik, {Six and seven loop Konishi from Luscher corrections},
  JHEP 1211 (2012) 002.
\newblock \href {http://arxiv.org/abs/1209.0791} {\path{arXiv:1209.0791}},
  \href {http://dx.doi.org/10.1007/JHEP11(2012)002}
  {\path{doi:10.1007/JHEP11(2012)002}}.

\bibitem{Arutyunov:2011mk}
G.~Arutyunov, S.~Frolov, S.~J. van Tongeren, {Bound States in the Mirror TBA},
  JHEP 1202 (2012) 014.
\newblock \href {http://arxiv.org/abs/1111.0564} {\path{arXiv:1111.0564}},
  \href {http://dx.doi.org/10.1007/JHEP02(2012)014}
  {\path{doi:10.1007/JHEP02(2012)014}}.

\bibitem{Arutyunov:2012tx}
G.~Arutyunov, S.~Frolov, A.~Sfondrini, {Exceptional Operators in N=4 super
  Yang-Mills}, JHEP 1209 (2012) 006.
\newblock \href {http://arxiv.org/abs/1205.6660} {\path{arXiv:1205.6660}},
  \href {http://dx.doi.org/10.1007/JHEP09(2012)006}
  {\path{doi:10.1007/JHEP09(2012)006}}.

\bibitem{Gromov:2010dy}
N.~Gromov, F.~Levkovich-Maslyuk, {Y-system and $\beta$-deformed N=4
  Super-Yang-Mills}, J.Phys. A44 (2011) 015402.
\newblock \href {http://arxiv.org/abs/1006.5438} {\path{arXiv:1006.5438}},
  \href {http://dx.doi.org/10.1088/1751-8113/44/1/015402}
  {\path{doi:10.1088/1751-8113/44/1/015402}}.

\bibitem{Arutyunov:2010gu}
G.~Arutyunov, M.~de~Leeuw, S.~J. van Tongeren, {Twisting the Mirror TBA}, JHEP
  1102 (2011) 025.
\newblock \href {http://arxiv.org/abs/1009.4118} {\path{arXiv:1009.4118}},
  \href {http://dx.doi.org/10.1007/JHEP02(2011)025}
  {\path{doi:10.1007/JHEP02(2011)025}}.

\bibitem{Bajnok:2010ud}
Z.~Bajnok, O.~el~Deeb, {6-loop anomalous dimension of a single impurity
  operator from AdS/CFT and multiple zeta values}, JHEP 1101 (2011) 054.
\newblock \href {http://arxiv.org/abs/1010.5606} {\path{arXiv:1010.5606}},
  \href {http://dx.doi.org/10.1007/JHEP01(2011)054}
  {\path{doi:10.1007/JHEP01(2011)054}}.

\bibitem{Ahn:2011xq}
C.~Ahn, Z.~Bajnok, D.~Bombardelli, R.~I. Nepomechie, {TBA, NLO Luscher
  correction, and double wrapping in twisted AdS/CFT}, JHEP 1112 (2011) 059.
\newblock \href {http://arxiv.org/abs/1108.4914} {\path{arXiv:1108.4914}},
  \href {http://dx.doi.org/10.1007/JHEP12(2011)059}
  {\path{doi:10.1007/JHEP12(2011)059}}.

\bibitem{deLeeuw:2012hp}
M.~de~Leeuw, S.~J. van Tongeren, {The spectral problem for strings on twisted
  $AdS_5 \times S^5$}, Nucl.Phys. B860 (2012) 339--376.
\newblock \href {http://arxiv.org/abs/1201.1451} {\path{arXiv:1201.1451}},
  \href {http://dx.doi.org/10.1016/j.nuclphysb.2012.03.004}
  {\path{doi:10.1016/j.nuclphysb.2012.03.004}}.

\bibitem{Arutyunov:2012zt}
G.~Arutyunov, M.~de~Leeuw, S.~J. van Tongeren, {The Quantum Deformed Mirror TBA
  I}, JHEP 1210 (2012) 090.
\newblock \href {http://arxiv.org/abs/1208.3478} {\path{arXiv:1208.3478}},
  \href {http://dx.doi.org/10.1007/JHEP10(2012)090}
  {\path{doi:10.1007/JHEP10(2012)090}}.

\bibitem{Arutyunov:2012ai}
G.~Arutyunov, M.~de~Leeuw, S.~J. van Tongeren, {The Quantum Deformed Mirror TBA
  II}, JHEP 1302 (2013) 012.
\newblock \href {http://arxiv.org/abs/1210.8185} {\path{arXiv:1210.8185}},
  \href {http://dx.doi.org/10.1007/JHEP02(2013)012}
  {\path{doi:10.1007/JHEP02(2013)012}}.

\bibitem{Correa:2012hh}
D.~Correa, J.~Maldacena, A.~Sever, {The quark anti-quark potential and the cusp
  anomalous dimension from a TBA equation}, JHEP 1208 (2012) 134.
\newblock \href {http://arxiv.org/abs/1203.1913} {\path{arXiv:1203.1913}},
  \href {http://dx.doi.org/10.1007/JHEP08(2012)134}
  {\path{doi:10.1007/JHEP08(2012)134}}.

\bibitem{Drukker:2012de}
N.~Drukker, {Integrable Wilson loops\hspace{2em}}\href
  {http://arxiv.org/abs/1203.1617} {\path{arXiv:1203.1617}}.

\bibitem{Gromov:2012eu}
N.~Gromov, A.~Sever, {Analytic Solution of Bremsstrahlung TBA}, JHEP 1211
  (2012) 075.
\newblock \href {http://arxiv.org/abs/1207.5489} {\path{arXiv:1207.5489}},
  \href {http://dx.doi.org/10.1007/JHEP11(2012)075}
  {\path{doi:10.1007/JHEP11(2012)075}}.

\bibitem{Basso:2011rs}
B.~Basso, {An exact slope for AdS/CFT\hspace{2em}}\href
  {http://arxiv.org/abs/1109.3154} {\path{arXiv:1109.3154}}.

\bibitem{Casteill:2007ct}
P.~Y. Casteill, C.~Kristjansen, {The Strong Coupling Limit of the Scaling
  Function from the Quantum String Bethe Ansatz}, Nucl. Phys. B785 (2007)
  1--18.
\newblock \href {http://arxiv.org/abs/0705.0890} {\path{arXiv:0705.0890}},
  \href {http://dx.doi.org/10.1016/j.nuclphysb.2007.06.011}
  {\path{doi:10.1016/j.nuclphysb.2007.06.011}}.

\bibitem{Gromov:2008en}
N.~Gromov, {Generalized Scaling Function at Strong Coupling}, JHEP 11 (2008)
  085.
\newblock \href {http://arxiv.org/abs/0805.4615} {\path{arXiv:0805.4615}},
  \href {http://dx.doi.org/10.1088/1126-6708/2008/11/085}
  {\path{doi:10.1088/1126-6708/2008/11/085}}.

\bibitem{Volin:2008kd}
D.~Volin, {The 2-loop generalized scaling function from the BES/FRS
  equation\,\,}\href {http://arxiv.org/abs/0812.4407} {\path{arXiv:0812.4407}}.

\bibitem{Basso:2007wd}
B.~Basso, G.~P. Korchemsky, J.~Kotanski, {Cusp anomalous dimension in maximally
  supersymmetric Yang- Mills theory at strong coupling}, Phys. Rev. Lett. 100
  (2008) 091601.
\newblock \href {http://arxiv.org/abs/0708.3933} {\path{arXiv:0708.3933}},
  \href {http://dx.doi.org/10.1103/PhysRevLett.100.091601}
  {\path{doi:10.1103/PhysRevLett.100.091601}}.

\bibitem{Kostov2008a}
I.~Kostov, D.~Serban, D.~Volin, {Functional BES equation}, JHEP 08 (2008) 101.
\newblock \href {http://arxiv.org/abs/0801.2542} {\path{arXiv:0801.2542}},
  \href {http://dx.doi.org/10.1088/1126-6708/2008/08/101}
  {\path{doi:10.1088/1126-6708/2008/08/101}}.

\bibitem{Hegedus:comm}
A.~Hegedus, {Private communication}.

\bibitem{Bajnok:comm}
Z.~Bajnok, {Private communication}.

\bibitem{Bombardelli:2009ns}
D.~Bombardelli, D.~Fioravanti, R.~Tateo, {Thermodynamic Bethe Ansatz for planar
  AdS/CFT: A Proposal}, J.Phys. A42 (2009) 375401.
\newblock \href {http://arxiv.org/abs/0902.3930} {\path{arXiv:0902.3930}},
  \href {http://dx.doi.org/10.1088/1751-8113/42/37/375401}
  {\path{doi:10.1088/1751-8113/42/37/375401}}.

\bibitem{GromovKKV}
N.~Gromov, V.~Kazakov, A.~Kozak, P.~Vieira, {Exact Spectrum of Anomalous
  Dimensions of Planar N = 4 Supersymmetric Yang-Mills Theory: TBA and excited
  states}, Lett. Math. Phys. 91 (2010) 265--287.
\newblock \href {http://arxiv.org/abs/0902.4458} {\path{arXiv:0902.4458}},
  \href {http://dx.doi.org/10.1007/s11005-010-0374-8}
  {\path{doi:10.1007/s11005-010-0374-8}}.

\bibitem{Arutyunov:2009ur}
G.~Arutyunov, S.~Frolov, {Thermodynamic Bethe Ansatz for the AdS$_5$xS$^5$
  Mirror Model}, JHEP 05 (2009) 068.
\newblock \href {http://arxiv.org/abs/0903.0141} {\path{arXiv:0903.0141}},
  \href {http://dx.doi.org/10.1088/1126-6708/2009/05/068}
  {\path{doi:10.1088/1126-6708/2009/05/068}}.

\bibitem{Cavaglia:2010nm}
A.~Cavaglia, D.~Fioravanti, R.~Tateo, {Extended Y-system for the $AdS_5/CFT_4$
  correspondence}, Nucl. Phys. B843 (2011) 302--343.
\newblock \href {http://arxiv.org/abs/1005.3016} {\path{arXiv:1005.3016}},
  \href {http://dx.doi.org/10.1016/j.nuclphysb.2010.09.015}
  {\path{doi:10.1016/j.nuclphysb.2010.09.015}}.

\bibitem{Balog:2011nm}
J.~Balog, A.~Hegedus, {$AdS_5\times S^5$ mirror TBA equations from Y-system and
  discontinuity relations}, JHEP 1108 (2011) 095.
\newblock \href {http://arxiv.org/abs/1104.4054} {\path{arXiv:1104.4054}},
  \href {http://dx.doi.org/10.1007/JHEP08(2011)095}
  {\path{doi:10.1007/JHEP08(2011)095}}.

\bibitem{Gromov:2011cx}
N.~Gromov, V.~Kazakov, S.~Leurent, D.~Volin, {Solving the AdS/CFT Y-system},
  JHEP 1207 (2012) 023.
\newblock \href {http://arxiv.org/abs/1110.0562} {\path{arXiv:1110.0562}},
  \href {http://dx.doi.org/10.1007/JHEP07(2012)023}
  {\path{doi:10.1007/JHEP07(2012)023}}.

\bibitem{Krichever:1996qd}
I.~Krichever, O.~Lipan, P.~Wiegmann, A.~Zabrodin, {Quantum integrable models
  and discrete classical Hirota equations}, Commun. Math. Phys. 188 (1997)
  267--304.
\newblock \href {http://arxiv.org/abs/hep-th/9604080}
  {\path{arXiv:hep-th/9604080}}, \href
  {http://dx.doi.org/10.1007/s002200050165} {\path{doi:10.1007/s002200050165}}.

\bibitem{Gromov:2010vb}
N.~Gromov, V.~Kazakov, Z.~Tsuboi, {PSU(2,2$|$4) Character of Quasiclassical
  AdS/CFT}, JHEP 07 (2010) 097.
\newblock \href {http://arxiv.org/abs/1002.3981} {\path{arXiv:1002.3981}},
  \href {http://dx.doi.org/10.1007/JHEP07(2010)097}
  {\path{doi:10.1007/JHEP07(2010)097}}.

\bibitem{Gromov:2010km}
N.~Gromov, V.~Kazakov, S.~Leurent, Z.~Tsuboi, {Wronskian Solution for AdS/CFT
  Y-system}, JHEP 1101 (2011) 155.
\newblock \href {http://arxiv.org/abs/1010.2720} {\path{arXiv:1010.2720}},
  \href {http://dx.doi.org/10.1007/JHEP01(2011)155}
  {\path{doi:10.1007/JHEP01(2011)155}}.

\bibitem{Tsuboi:2009ud}
Z.~Tsuboi, {Solutions of the T-system and Baxter equations for supersymmetric
  spin chains}, Nucl.Phys. B826 (2010) 399--455.
\newblock \href {http://arxiv.org/abs/0906.2039} {\path{arXiv:0906.2039}},
  \href {http://dx.doi.org/10.1016/j.nuclphysb.2009.08.009}
  {\path{doi:10.1016/j.nuclphysb.2009.08.009}}.

\bibitem{Tsuboi:2011iz}
Z.~Tsuboi, {Wronskian solutions of the T, Q and Y-systems related to infinite
  dimensional unitarizable modules of the general linear superalgebra
  gl(M$|$N)}, Nucl.Phys. B870 (2013) 92--137.
\newblock \href {http://arxiv.org/abs/1109.5524} {\path{arXiv:1109.5524}},
  \href {http://dx.doi.org/10.1016/j.nuclphysb.2013.01.007}
  {\path{doi:10.1016/j.nuclphysb.2013.01.007}}.

\bibitem{Volin:2010xz}
D.~Volin, {String hypothesis for gl(n$|$m) spin chains: a particle/hole
  democracy}, Lett.Math.Phys. 102 (2012) 1--29.
\newblock \href {http://arxiv.org/abs/1012.3454} {\path{arXiv:1012.3454}},
  \href {http://dx.doi.org/10.1007/s11005-012-0570-9}
  {\path{doi:10.1007/s11005-012-0570-9}}.

\bibitem{GKLV:wip}
N.~Gromov, V.~Kazakov, S.~Leurent, D.~Volin, {Work in progress}.

\bibitem{Zudilin}
W.~Zudilin, {Algebraic relations for multiple zeta values}, Uspekhi Mat. Nauk
  58~(1) (2003) 3--32.

\bibitem{WaldschmidtIntro}
M.~Waldschmidt, {Valeurs z\^eta multiples. Une introduction}, Journal de
  th\'eorie des nombres de Bordeaux 12~(4) (2000) 581--595.
\newblock \href {http://dx.doi.org/10.5802/jtnb.298}
  {\path{doi:10.5802/jtnb.298}}.

\bibitem{Ablinger:2013cf}
J.~Ablinger, J.~Blumlein, C.~Schneider, {Analytic and Algorithmic Aspects of
  Generalized Harmonic Sums and Polylogarithms\,\,}\href
  {http://arxiv.org/abs/1302.0378} {\path{arXiv:1302.0378}}.

\bibitem{Goncharov}
A.~Goncharov, {Multiple polylogarithms, cyclotomy and modular complexes},
  Mathematical Research Letters 5 (1998) 497Ð516.

\bibitem{Hoffman1997477}
M.~E. Hoffman, The algebra of multiple harmonic series, Journal of Algebra
  194~(2) (1997) 477 -- 495.
\newblock \href {http://dx.doi.org/10.1006/jabr.1997.7127}
  {\path{doi:10.1006/jabr.1997.7127}}.

\bibitem{BaileyPSLQ}
D.~H. Bailey, D.~J. Broadhurst, {Parallel integer relation detection:
  Techniques and applications}, Math. Comput. 70~(236) (2001) 1719--1736.

\bibitem{Broadhurst:1996kc}
D.~J. Broadhurst, D.~Kreimer, {Association of multiple zeta values with
  positive knots via Feynman diagrams up to 9 loops}, Phys.Lett. B393 (1997)
  403--412.
\newblock \href {http://arxiv.org/abs/hep-th/9609128}
  {\path{arXiv:hep-th/9609128}}, \href
  {http://dx.doi.org/10.1016/S0370-2693(96)01623-1}
  {\path{doi:10.1016/S0370-2693(96)01623-1}}.

\bibitem{Ohno199939}
Y.~Ohno, A generalization of the duality and sum formulas on the multiple zeta
  values, Journal of Number Theory 74~(1) (1999) 39 -- 43.
\newblock \href {http://dx.doi.org/10.1006/jnth.1998.2314}
  {\path{doi:10.1006/jnth.1998.2314}}.

\bibitem{HoffmanOhno}
M.~E. Hoffman, Y.~Ohno, Relations of multiple zeta values and their algebraic
  expression, J. Algebra 262 (2003) 332--347.
\newblock \href {http://arxiv.org/abs/math/0010140}
  {\path{arXiv:math/0010140}}, \href
  {http://dx.doi.org/http://dx.doi.org/10.1016/S0021-8693(03)00016-4}
  {\path{doi:http://dx.doi.org/10.1016/S0021-8693(03)00016-4}}.

\bibitem{CambridgeJournals:414263}
K.~Ihara, M.~Kaneko, D.~Zagier, Derivation and double shuffle relations for
  multiple zeta values, Compositio Mathematica 142  307--338.
\newblock \href {http://dx.doi.org/10.1112/S0010437X0500182X}
  {\path{doi:10.1112/S0010437X0500182X}}.

\bibitem{Blumlein:2009cf}
J.~Blumlein, D.~Broadhurst, J.~Vermaseren, {The Multiple Zeta Value Data Mine},
  Comput.Phys.Commun. 181 (2010) 582--625.
\newblock \href {http://arxiv.org/abs/0907.2557} {\path{arXiv:0907.2557}},
  \href {http://dx.doi.org/10.1016/j.cpc.2009.11.007}
  {\path{doi:10.1016/j.cpc.2009.11.007}}.

\bibitem{Ablinger:2011te}
J.~Ablinger, J.~Blumlein, C.~Schneider, {Harmonic Sums and Polylogarithms
  Generated by Cyclotomic Polynomials}, J.Math.Phys. 52 (2011) 102301.
\newblock \href {http://arxiv.org/abs/1105.6063} {\path{arXiv:1105.6063}},
  \href {http://dx.doi.org/10.1063/1.3629472} {\path{doi:10.1063/1.3629472}}.

\bibitem{Gromov:2008gj}
N.~Gromov, V.~Kazakov, P.~Vieira, {Finite Volume Spectrum of 2D Field Theories
  from Hirota Dynamics}, JHEP 0912 (2009) 060.
\newblock \href {http://arxiv.org/abs/0812.5091} {\path{arXiv:0812.5091}},
  \href {http://dx.doi.org/10.1088/1126-6708/2009/12/060}
  {\path{doi:10.1088/1126-6708/2009/12/060}}.

\bibitem{Kazakov:2010kf}
V.~Kazakov, S.~Leurent, {Finite Size Spectrum of SU(N) Principal Chiral Field
  from Discrete Hirota Dynamics\,\,}\href {http://arxiv.org/abs/1007.1770}
  {\path{arXiv:1007.1770}}.

\bibitem{Janik:2006dc}
R.~A. Janik, {The AdS$_5$xS$^5$ superstring worldsheet S-matrix and crossing
  symmetry}, Phys. Rev. D73 (2006) 086006.
\newblock \href {http://arxiv.org/abs/hep-th/0603038}
  {\path{arXiv:hep-th/0603038}}, \href
  {http://dx.doi.org/10.1103/PhysRevD.73.086006}
  {\path{doi:10.1103/PhysRevD.73.086006}}.

\bibitem{Volin:2009uv}
D.~Volin, {Minimal solution of the AdS/CFT crossing equation}, J.Phys. A42
  (2009) 372001.
\newblock \href {http://arxiv.org/abs/0904.4929} {\path{arXiv:0904.4929}},
  \href {http://dx.doi.org/10.1088/1751-8113/42/37/372001}
  {\path{doi:10.1088/1751-8113/42/37/372001}}.

\bibitem{Gromov:comm}
N.~Gromov, {Private communication}.

\bibitem{Schnetz}
O.~Schnetz,
  \href{http://www.raumzeitmaterie.de/veranstaltungen.php?evt=select\&sqn=8970%
\&lang=en}{{Single-valued multiple-zeta-values}}, {SFB Colloquium \emph{Novel
  Methods for Perturbative QFT}, Humboldt-Universit\"at zu Berlin, May 2013}.
\newline\urlprefix\url{http://www.raumzeitmaterie.de/veranstaltungen.php?evt=s%
elect\&sqn=8970\&lang=en}

\bibitem{Broadhurst:1995km}
D.~J. Broadhurst, D.~Kreimer, {Knots and numbers in $\phi^4$ theory to 7 loops
  and beyond}, Int.J.Mod.Phys. C6 (1995) 519--524.
\newblock \href {http://arxiv.org/abs/hep-ph/9504352}
  {\path{arXiv:hep-ph/9504352}}, \href
  {http://dx.doi.org/10.1142/S012918319500037X}
  {\path{doi:10.1142/S012918319500037X}}.

\bibitem{Brown:2010bw}
F.~Brown, O.~Schnetz, {A K3 in $\phi^4$\,\,}, Duke Mathematical Journal
  161~(10) (2012) 1817--1862.
\newblock \href {http://arxiv.org/abs/1006.4064} {\path{arXiv:1006.4064}},
  \href {http://dx.doi.org/10.1090/conm/538/10594}
  {\path{doi:10.1090/conm/538/10594}}.

\bibitem{Dorey:1996re}
P.~Dorey, R.~Tateo, {Excited states by analytic continuation of TBA equations},
  Nucl. Phys. B482 (1996) 639--659.
\newblock \href {http://arxiv.org/abs/hep-th/9607167}
  {\path{arXiv:hep-th/9607167}}, \href
  {http://dx.doi.org/10.1016/S0550-3213(96)00516-0}
  {\path{doi:10.1016/S0550-3213(96)00516-0}}.

\end{thebibliography}

\end{document}